\documentclass[a4paper]{article}

\usepackage[widelayout,hyperrefblack,copyright]{research17}
\usepackage{verbatim}

\usepackage{pgfplots}
\pgfplotsset{compat=1.14}
\usepgfplotslibrary{units}
\usetikzlibrary{spy,backgrounds}
\usetikzlibrary{decorations.markings}
\usepackage{pgfplotstable}

\usepackage[format=hang]{caption}

\renewcommand{\mycopyright}{Moser, Wang, Wigger, submitted, 26 September 2017}

\newcommand{\Vmax}{\const{V}_{\textnormal{max}}}

\newcommand{\C}{\const{C}}
\renewcommand{\indicatorfunction}{\mathop{}\!\constb{I}}
\newcommand{\Xs}{\bar{X}}

\begin{document}

\title{Capacity Results on Multiple-Input Single-Output
  Wireless Optical Channels}

\author{Stefan M.~Moser, Ligong Wang, Michèle Wigger\thanks{S.~Moser
    is with the Signal and Information Processing Lab, ETH Zurich,
    Switzerland and with the Department of Electrical and Computer
    Engineering at National Chiao Tung University (NCTU), Hsinchu,
    Taiwan. L. Wang is with ETIS---Université Paris Seine, Université
    de Cergy-Pontoise, ENSEA, CNRS, Cergy-Pontoise, France. M. Wigger
    is with LTCI, Telecom ParisTech, Université Paris-Saclay, 75013
    Paris, France.
    This work was presented in parts at the \emph{2017 IEEE
      International Symposium on Information Theory (ISIT)}, Aachen,
    Germany, Jun.~25--30, 2017 and will be presented in parts at the
    \emph{2017 IEEE Information Theory Workshop (ITW)}, Kaohsiung,
    Taiwan, Nov.~6--10, 2017.}}

\date{26 September 2017}

\maketitle

\begin{abstract}
  This paper derives upper and lower bounds on the capacity of the
  multiple-input single-output free-space optical intensity channel
  with signal-independent additive Gaussian noise subject to both an
  average-intensity and a peak-intensity constraint. In the limit
  where the signal-to-noise ratio (SNR) tends to infinity, the
  asymptotic capacity is specified, while in the limit where the SNR
  tends to zero, the exact slope of the capacity is also given.
\end{abstract}

\vspace{0.5cm}
\noindent
\textbf{Index Terms} --- Average- and peak-power constraint, channel
capacity, direct detection, Gaussian noise, infrared communication,
multiple-input single-output (MISO) channel, optical communication.

\section{Introduction}
\label{sec:introduction}

Optical wireless communication is a form of communication in which
visible, infrared, or ultraviolet light is transmitted in free space
(air or vacuum) to carry a message to its destination. Recent works
suggest that it is a promising solution to replacing some of the
existing radio-frequency (RF) wireless communication systems in order
to prevent future rate bottlenecks \cite{kahnbarry97_1,
  hranilovic05_1, khalighiuysal14_1}. Particularly attractive are
simple \emph{intensity-modulation--direct-detection (IM-DD)}
systems. In such a system, the transmitter modulates the intensity of
optical signals coming from light emitting diodes (LEDs) or laser
diodes (LDs), and the receiver measures incoming optical intensities
by means of photodetectors. The electrical output signals of the
photodetectors are essentially proportional to the incoming optical
intensities, but are corrupted by thermal noise of the photodetectors,
relative-intensity noise of random intensity fluctuations inherent to
low-cost LEDs and LDs, and shot noise caused by ambient light. In a
first approximation, noise coming from these sources is usually
modeled as being additive Gaussian and independent of the transmitted
light signal; see \cite{kahnbarry97_1, hranilovic05_1}.
 
The free-space optical intensity channel has been extensively studied
in the literature. In the \emph{single-input single-output (SISO)}
scenario, where the transmitter employs a single transmit LED or LD,
and the receiver a single photodetector, the works \cite{wigger03_1,
  lapidothmoserwigger09_7} established upper and lower bounds on the
capacity of this channel that are asymptotically tight in both
high-signal-to-noise-ratio (SNR) and low-SNR limits. Improved bounds
at finite SNR have subsequently been presented in \cite{mckellips04_1,
  faridhranilovic09_1, faridhranilovic10_1,
  thangarajkramerbocherer17_1}. For the \emph{multiple-input and
  multiple\?output (MIMO)} optical intensity channel, where the
transmitter is equipped with multiple LEDs or LDs, and the receiver
with multiple photodetectors, the recent work
\cite{mosermylonakiswangwigger17_1} determined the asymptotic capacity
in the high-SNR limit when \emph{the channel matrix is of full column
  rank}. A special case of this MIMO result was independently solved
in \cite{chaabanrezkialouini17_2}.

Previous to \cite{mosermylonakiswangwigger17_1,
  chaabanrezkialouini17_2}, various code constructions for this setup
have been proposed in \cite{haasshapirotarokh02_1, bayakischober10_1,
  songcheng13_1, mrouehbelfiore17_1app}. When there is no crosstalk so
the MIMO channel can be modeled through a diagonal channel matrix,
bounds on capacity were presented in
\cite{thangarajkramerbocherer17_1, chaabanrezkialouini17_1}.


The current work is concerned with the \emph{multiple-input and
  single-output (MISO)} channel. (Clearly, the channel matrix of the
MISO channel cannot have full column rank.) Our main results, some of
which presented in part in
\cite{mosermylonakiswangwigger17_1,moserwangwigger17_1app}, include
\begin{itemize}
\item several upper and lower bounds on the capacity;
\item high-SNR asymptotic capacity for all parameter ranges; and
\item low-SNR capacity slope in terms of a maximum variance on the
  input.
\end{itemize}

The high-SNR asymptotic capacity is proven based on two capacity lower
bounds (Propositions~\ref{prop:lower} and~\ref{prop:lower2}) derived
using the Entropy Power Inequality (EPI), and an upper bound
(Proposition~\ref{prop:finiteSNR}) derived using the duality technique
\cite{lapidothmoser03_3}. These proof techniques are similar to those
in in \cite{lapidothmoserwigger09_7, mosermylonakiswangwigger17_1,
  thangarajkramerbocherer17_1}, but also involve some nontrivial
optimization. In particular, the optimal input distribution at high
SNR involves LED cooperation (compared to independent signaling in the
MIMO full-column-rank case \cite{mosermylonakiswangwigger17_1}), and,
with certain probabilities, assigns to each LED a truncated
exponential distribution, whose parameters must be carefully chosen.

The low-SNR capacity slope is proven using a simple upper bound
(Proposition~\ref{prop:simpler_upper}) and a classic asymptotic lower
bound by Prelov and Van der Meulen \cite{prelovvandermeulen93_1}. The
expression of this slope involves a maximization of variance, which
can be easily computed numerically (Lemma~\ref{lem:maximum_variance}).

All the above-mentioned bounds (except the asymptotic one from
\cite{prelovvandermeulen93_1}) hold at all SNR, and not only in the
high- or low-SNR limits. They are compared numerically, together with
another indirect upper bound (Proposition~\ref{prp:SISO}), which
utilizes upper bounds on the capacity of the SISO channel from
\cite{lapidothmoserwigger09_7, mckellips04_1,
  thangarajkramerbocherer17_1}. 

The remainder of this paper is structured as follows. After a few
remarks on notation, Section~\ref{sec:channel-model} lays out the
specific details of the investigated channel model.
Section~\ref{sec:equiv-capac-form} reviews the known capacity results
from \cite{mosermylonakiswangwigger17_1} and proves a fundamental
proposition giving the optimal structure of an input to the MISO
channel with both active average- and peak-power
constraints. Section~\ref{sec:bounds-capac-alpha} then presents all
new upper and lower bounds on the capacity and gives the correct
high-SNR and low-SNR asymptotics. The detailed proofs for the lower
bounds can be found in Section~\ref{sec:lower-bound} and the proof for
one of the upper bounds in
Section~\ref{sec:finiteSNR}. Section~\ref{sec:asymptotic} shows the
analysis of the high-SNR capacity. The paper is concluded in
Section~\ref{sec:concluding-remarks}.

We meticulously distinguish between random and deterministic
quantities. A random variable is denoted by a capital Roman letter,
e.g., $Z$, while its realization is denoted by the corresponding small
Roman letter, e.g., $z$. Vectors are bold-faced, e.g., $\vect{X}$
denotes a random vector and $\vect{x}$ its realization. Constants are
typeset either in small Romans, in Greek letters or in a special font,
e.g., $\EE$ or $\amp$. Entropy is typeset as $\HH(\cdot)$,
differential entropy as $\hh(\cdot)$, and $\II(\cdot;\cdot)$ denotes
the mutual information \cite{shannon48_1}. The logarithm function
$\log(\cdot)$ denotes the natural logarithm.

\section{Channel Model}
\label{sec:channel-model}

Consider a communication link where the transmitter is equipped with
$\nt$ LEDs (or LDs), $\nt\ge 2$, and the receiver with a single
photodetector.  The photodetector receives a superposition of the
signals sent by the LEDs, and we assume that the crosstalk between
LEDs is constant. Hence, the channel output is given by
\begin{IEEEeqnarray}{c}
  Y = \trans{\vect{h}} \vect{x} + Z,
  \label{eq:channel}
\end{IEEEeqnarray}
where the $\nt$-vector $\vect{x}$ denotes the channel input, whose
entries are proportional to the optical intensities of the
corresponding LEDs, and are therefore nonnegative:
\begin{IEEEeqnarray}{c}
  X_k\in \Reals_0^+, \quad k=1, \ldots, \nt;
  \label{eq:nonzero}
\end{IEEEeqnarray}
where the length-$\nt$ row vector $\trans{\vect{h}}$ is the constant
channel state vector with nonnegative entries, which, without loss of
generality, we assume to be ordered:
\begin{IEEEeqnarray}{c}
  h_1\ge h_2\ge \cdots \ge h_{\nt} > 0;
\end{IEEEeqnarray}
and where $Z\sim \Normal{0}{\sigma^2}$ is additive Gaussian noise.
Note that, in contrast to the input $\vect{x}$, the output $Y$ can be
negative.

Inputs are subject to a peak-power (peak-intensity) and an
average-power (average-intensity) constraint:
\begin{IEEEeqnarray}{rCl}
  \bigPrv{X_k>\amp} & = & 0, \quad \forall\,k\in\{1, \ldots, \nt\},
  \label{eq:peak}
  \\
  \sum_{k=1}^{\nt} \bigE{X_k} & \le & \EE,
  \label{eq:average}
\end{IEEEeqnarray}
for some fixed parameters $\amp,\EE>0$. 
Note that the average-power constraint is on the expectation of the
channel input and not on its square. Also note that $\amp$ describes
the maximum power of each single LED, while $\EE$ describes the
allowed average total power of all LEDs together.

We denote the ratio between the allowed average power and the allowed
peak power by $\alpha$:
\begin{IEEEeqnarray}{c}
  \alpha \eqdef \frac{\EE}{\amp},
  \label{eq:2}
\end{IEEEeqnarray}
where $0 < \alpha \le \nt$. For $\alpha=\nt$ the average-power
constraint is inactive in the sense that it is automatically satisfied
whenever the peak-power constraint is satisfied. Thus, $\alpha=\nt$
corresponds to the case with only a peak-power constraint. 

We denote the capacity of the channel \eqref{eq:channel} with allowed
peak power $\amp$ and allowed average power $\EE$ by
$\C_{\trans{\vect{h}},\sigma^2}(\amp,\EE)$. The capacity is given by
\cite{shannon48_1}
\begin{IEEEeqnarray}{c}
  \C_{\trans{\vect{h}},\sigma^2}(\amp,\EE)= \sup_{Q_{\vect{X}}}
  \II(\vect{X}; Y) 
  \label{eq:C}
\end{IEEEeqnarray}
where the supremum is over all laws $Q_{\vect{X}}$ on $\vect{X}$
satisfying \eqref{eq:nonzero}, \eqref{eq:peak}, and
\eqref{eq:average}.  When only an average-power constraint is imposed,
capacity is denoted by $\C_{\trans{\vect{h}},\sigma^2}(\EE)$. It is
given as in \eqref{eq:C} except that the supremum is taken over all
laws $Q_{\vect{X}}$ on $\vect{X}$ satisfying \eqref{eq:nonzero} and
\eqref{eq:average}.

\section{Equivalent Capacity Formulas}
\label{sec:equiv-capac-form}

Denote 
\begin{IEEEeqnarray}{rCl}
  \subnumberinglabel{eq:s}
  s_0 & \eqdef & 0
  \\
  s_k & \eqdef & \sum_{k'=1}^{k} h_{k'}, \qquad k\in\{1,\ldots, \nt\},
  \\
  \noalign{\noindent and\vspace{\jot}}
  \IEEEnosubnumber
  \Xs & \eqdef & \trans{\vect{h}} \vect{X}=\sum_{k=1}^{\nt} h_k X_k.
  \label{eq:Xs}
\end{IEEEeqnarray}
Also, define the random variable $U$ over the alphabet
$\{1,\ldots, \nt\}$ to indicate in which interval $\Xs$ lies:
\begin{IEEEeqnarray}{rCl}
  \subnumberinglabel{eq:U}
  \Bigl(U=1\Bigr)  & \iff & \Bigl(\Xs \in [\amp s_{0}, \amp s_1]\Bigr)
  \label{eq:Ua}
  \\
  \noalign{\noindent and for $k\in\{2,\ldots, \nt\}$:\vspace{\jot}}
  \Bigl(U = k\Bigr) & \iff & \Bigl(\Xs \in (\amp s_{k-1}, \amp
  s_k]\Bigr). 
  \label{eq:Ub}
\end{IEEEeqnarray}

Now notice that because $\vect{X}\markov \Xs \markov Y$ form a Markov
chain and because $\Xs$ is a function of $\vect{X}$, we have
\begin{IEEEeqnarray}{c}
  \II(\vect{X}; Y) = \II(\Xs;Y).
  \label{eq:MI_equivalence}
\end{IEEEeqnarray}
Hence the MISO channel \eqref{eq:channel} is equivalent to a SISO
channel with input $\Xs$ and output $Y=\Xs+Z$ with the
power-constraints \eqref{eq:peak} and \eqref{eq:average} on $\vect{X}$
transformed to a set of admissible distributions for $\Xs$.  So,
\begin{IEEEeqnarray}{rCl}
  \C_{\trans{\vect{h}},\sigma^2}(\amp,\EE)
  & = & \max_{Q_{\Xs}} \II(\Xs;Y), 
  \label{eq:equiv}
\end{IEEEeqnarray}
where $Q_{\Xs}$ is restricted to the set of admissible
distributions. Characterizing this set of admissible distributions is
relatively straightforward when there is only an average- or only a
peak-power constraint, but is more involved in general.  This is the
subject of the following three propositions.

\sloppy
If $\vect{X}$ is only subject to an average-power constraint~$\EE$
(and no peak-power constraint), then $\Xs$ is only subject to an
average-power constraint $h_{1} \EE$.
\begin{proposition}[Only Average-Power Constraint]
  \label{prop:MISOaverageonly}
  Without a  peak-power constraint,
  \begin{IEEEeqnarray}{c}
    \label{eq:MISOaverageonly}
    \C_{\trans{\vect{h}},\sigma^2} (\EE)
    = \max_{\substack{Q_{\Xs}\colon \Xs \in[0,\infty),\\
        \qquad \eE{\Xs} \leq h_1 \EE}} \II(\Xs;Y)
    = \C_{1,\sigma^2}(h_1 \EE),
  \end{IEEEeqnarray}
  where $\C_{1,\sigma^2}(h_1 \EE)$ denotes the capacity of a SISO
  channel with unit channel gain under average-power constraint
  $h_1 \EE$.
\end{proposition}
\fussy
\begin{IEEEproof}
  When $\vect{X}$ satisfies \eqref{eq:average}, we have
  \begin{IEEEeqnarray}{c}
    \E{\Xs} = \sum_{k=1}^{\nt} h_{k} \E{X_k} \le  h_1\EE, 
  \end{IEEEeqnarray}
  so $\C_{\trans{\vect{h}},\sigma^2} (\EE) \le \C_{1,\sigma^2}(h_1
  \EE)$.  For the reverse direction, to achieve any target
  distribution on $\Xs$ satisfying $\E{\Xs} \le h_1 \EE$, the
  transmitter can let the LED corresponding to $h_1$ send $\Xs/h_1$
  and all the other LEDs send zero.
\end{IEEEproof}
\bigskip

When $\vect{X}$ is only subject to a peak-power constraint~$\amp$ (and
no average-power constraint), then $\Xs$ is only subject to a
peak-power constraint $s_{\nt} \amp$.  Moreover, for
$\alpha \ge \frac{\nt}{2}$, the average-power constraint is inactive
because the capacity-achieving input distribution in the absence of a
peak-power constraint can be shown to be symmetric
around~$\frac{\amp}{2}$ \cite[Prop.~1]{mosermylonakiswangwigger17_1}.
\begin{proposition}[Only Peak-Power Constraint is Active]
  \label{prop:MISOpeakonly}
  When $\alpha \ge \frac{\nt}{2}$,
  \begin{IEEEeqnarray}{c}
    \C_{\trans{\vect{h}},\sigma^2} (\amp,\alpha \amp)
    = \max_{Q_{\Xs}\colon \Xs \in[0, s_{\nt} \amp]} \II(\Xs;Y)  
    = \C_{1,\sigma^2}\left( s_{\nt}\amp,\frac{ s_{\nt}\amp}{2}
    \right),
    \label{eq:MISOpeakonly}
  \end{IEEEeqnarray}
  where
  $\C_{1,\sigma^2}\left( s_{\nt}\amp,\frac{ s_{\nt}\amp}{2} \right)$
  denotes the capacity of a SISO channel with unit channel gain under
  peak-power constraint $s_{\nt}\amp$ and average-power constraint
  $\frac{ s_{\nt}\amp}{2}$.
\end{proposition}

\begin{IEEEproof}
  When $\vect{X}$ satisfies the peak-power constraint \eqref{eq:peak},
  $\Xs$ must satisfy $\Xs\le s_{\nt} \amp$ with probability one. Hence
  $\C_{\trans{\vect{h}},\sigma^2} (\amp,\alpha \amp)$ cannot exceed
  the capacity of the SISO channel with allowed peak power
  $s_{\nt} \amp$. By \cite[Prop.~9]{lapidothmoserwigger09_7}, for a
  SISO channel with allowed peak power $s_{\nt} \amp$, adding an
  average-power constraint of $\frac{s_{\nt} \amp}{2}$ does not affect
  its capacity. We hence know that the left-hand side (LHS) of
  \eqref{eq:MISOpeakonly} is upper-bounded by its right-hand side
  (RHS).
	
  For the reverse direction, consider any target distribution on $\Xs$
  satisfying peak-power constraint $s_{\nt} \amp$ and average-power
  constraint $\frac{1}{2}s_{\nt} \amp$. We need only to show that such
  an $\Xs$ can be generated by some distribution for $\vect{X}$
  satisfying peak- and average-power constraints $\amp$ and
  $\alpha \amp$, respectively. To this end, we let the transmitter
  send the same signal on all LEDs:
  \begin{IEEEeqnarray}{c}
    X_k = \frac{\Xs}{s_{\nt}},\quad k\in\{1,\ldots,\nt\}.
  \end{IEEEeqnarray}
  One can easily check that both constraints are indeed satisfied by
  this choice.
\end{IEEEproof}
\bigskip

As already mentioned, describing the set of admissible distributions
is more complicated when $\alpha < \frac{\nt}{2}$. Recall the
definition of $U$ in \eqref{eq:U} and let $p_k\eqdef \Prv{U=k}$ for
$k=1,\ldots, \nt$.
 
\begin{proposition}[Active Average- and Peak-Power Constraints]
  \label{prop:equivalence}
  When $\alpha < \frac{\nt}{2}$, 
  \begin{IEEEeqnarray}{rCl}
    \label{eq:C_equiv}
    \C_{\trans{\vect{h}},\sigma^2}(\amp,\alpha \amp)
    & = & \max_{Q_{\Xs}} \II(\Xs;Y) 
  \end{IEEEeqnarray}
  where the maximization is over all laws on $\Xs\in \Reals_0^+$
  satisfying
  \begin{IEEEeqnarray}{c}
    \subnumberinglabel{eq:newconstraints}
    \Prv{\Xs > s_{\nt} \amp} = 0
    \label{eq:newconstraintsa}
  \end{IEEEeqnarray}
  and
  \begin{IEEEeqnarray}{c}
    \IEEEyessubnumber*
    \sum_{k=1}^{\nt} p_{k} \left( \frac{\eEcond{\Xs}{U=k} - \amp
        s_{k-1}}{h_{k}} +(k-1)\amp \right)
    \leq \alpha \amp. 
    \label{eq:newconstraintsb}
  \end{IEEEeqnarray}
\end{proposition}

The proof of Proposition~\ref{prop:equivalence} is based on the
following lemma, which will be of further interest in this paper.

\begin{lemma}
  \label{lem:energy_efficient}
  Without loss in optimality, the maximization in \eqref{eq:C} can be
  restricted to distributions $Q_{\vect{X}}$ of the input vector
  $\vect{X}$ satisfying for all $k\in\{1,\ldots,\nt\}$, with
  probability one,
  \begin{IEEEeqnarray}{c}
    \label{eq:lemma_cond}
    \Bigl(X_k>0\Bigr) \implies \Bigl(X_{1} = \cdots = X_{k-1} =
    \amp\Bigr). 
  \end{IEEEeqnarray}
\end{lemma}
\begin{IEEEproof}
  Fix $X_1,\ldots, X_{\nt}$ satisfying the peak- and average-power
  constraints \eqref{eq:peak} and \eqref{eq:average} and let $\Xs$
  and $U$ be defined as in \eqref{eq:Xs} and \eqref{eq:U}.  Define
  also a set of new inputs $X_1^*, \ldots, X_{\nt}^*$ that with
  probability $p_k=\Prv{U=k}$ take on values
  \begin{IEEEeqnarray}{rCl}
    \subnumberinglabel{eq:energy-efficient}
    X_1^* = \cdots = X_{k-1}^*
    & = & \amp
    \\
    X_k^* & = & \frac{\Xs-\amp s_{k-1}}{h_{k}}\bigg|_{U=k}
    \\
    X_{k+1}^* = \cdots = X_{\nt}^* & = & 0.
  \end{IEEEeqnarray}
  Notice that
  \begin{IEEEeqnarray}{rCl}
    \Xs^* & \eqdef & \sum_{k=1}^{\nt} h_k X_k^*
    \\
    & = & \sum_{k'=1}^{\nt} p_{k'} \sum_{k=1}^{\nt}
    h_k X_k^*\bigg|_{U=k'}
    \\
    & = & \sum_{k'=1}^{\nt} p_{k'} \left( \sum_{k=1}^{k'-1}
      h_k \amp +  h_{k'} \frac{\Xs\big|_{U=k'}-\amp s_{k'-1}}{h_{k'}}
    \right) 
    \\
    & = & \sum_{k'=1}^{\nt} p_{k'} \Xs\big|_{U=k'}
    \\
    & = & \Xs
  \end{IEEEeqnarray}
  and hence by \eqref{eq:MI_equivalence}
  \begin{IEEEeqnarray}{c}
    \II(\vect{X}^*;Y) = \II(\Xs^*;Y) = \II(\Xs;Y) = \II(\vect{X};Y).
  \end{IEEEeqnarray}
  Moreover, the new inputs $X_1^*, \ldots, X_{\nt}^*$ are admissible
  because they trivially satisfy the peak-power constraint
  \eqref{eq:peak} and their average power does not exceed the average
  power of the original inputs $X_1,\ldots, X_{\nt}$:
  \begin{IEEEeqnarray}{c}
    \sum_{k=1}^{\nt} \E{X_k^*}  \leq \sum_{k=1}^{\nt} \E{X_k}.
  \end{IEEEeqnarray}
  In fact, it is not hard to see that among all input assignments
  generating a sum-input $\Xs \in (s_{k-1}\amp,s_k\amp]$, the choice
  in \eqref{eq:energy-efficient} consumes least input energy
  $\sum_{k=1}^{\nt} \E{X_k^*}$.
\end{IEEEproof}
\bigskip

\begin{IEEEproof}[Proof of Proposition~\ref{prop:equivalence}]
  By Lemma~\ref{lem:energy_efficient}, we can restrict attention to
  distributions on $\vect{X}$ satisfying the implication in
  \eqref{eq:lemma_cond}. Notice that for these distributions there is
  a one-to-one correspondence between $\vect{X}$ and $\Xs$.
  The average input power $\E{\sum_{k=1}^{\nt}X_k}$ can thus entirely
  be expressed in terms of $\Xs$. In fact, since the input power
  consumed for $\Xs\in(s_{k-1} \amp, s_k \amp]$ is
  \begin{IEEEeqnarray}{c}
    \sum_{k'=1}^{\nt} X_k' =\frac{\Xs - \amp s_{k-1}}{h_k} +
    (k-1)\amp
    \qquad \textnormal{(conditional on $\Xs\in(s_{k-1} \amp,  s_k
      \amp]$)}, 
  \end{IEEEeqnarray}  
  the average input power is
  \begin{IEEEeqnarray}{rCl}
    \E{\sum_{k'=1}^{\nt} X_{k'}}
    & = & \sum_{k=1}^{\nt}p_k \Econd{\sum_{k'=1}^{\nt} X_{k'}}{U=k}
    \\[1mm]
    & = & \sum_{k=1}^{\nt} p_k \left(\frac{\eEcond{\Xs}{U=k} - \amp
        s_{k-1}}{h_{k}} +(k-1)\amp\right).
    \label{eq:m30}
  \end{IEEEeqnarray}
  The proposition now follows from \eqref{eq:equiv} and \eqref{eq:m30}
  and because $\Xs \in [0, s_{\nt} \amp]$.
\end{IEEEproof}
\bigskip

Bounds on the capacities $\C_{1,\sigma^2}( h_1 \EE)$ and
$\C_{1,\sigma^2}\left( s_{\nt}\amp,\frac{ s_{\nt}\amp}{2} \right)$
were presented in \cite{lapidothmoserwigger09_7, smith71_1,
  mckellips04_1, thangarajkramerbocherer17_1}. Moreover,
\cite{lapidothmoserwigger09_7} also characterizes exactly the high-SNR
asymptotic behavior of these two capacities and the low-SNR asymptotic
behavior of
$\C_{1,\sigma^2}\left( s_{\nt}\amp,\frac{ s_{\nt}\amp}{2} \right)$.
In the rest of this paper we focus on the case with active average-
and peak-power constraints, so $\alpha < \frac{\nt}{2}$, and we bound
the RHS of \eqref{eq:C_equiv} under constraints
\eqref{eq:newconstraints}.

\section{Bounds on  Capacity  When \texorpdfstring{$\alpha <  \frac{\nt}{2}$}{alpha < nt/2}}
\label{sec:bounds-capac-alpha}

Throughout this section we assume $\alpha < \frac{\nt}{2}$. Some of
our bounds depend on whether $\alpha$ is larger or smaller than the
threshold
\begin{IEEEeqnarray}{c}
  \label{eq:alpha48}
  \alpha_{\textnormal{th}} \eqdef 
  \frac{1}{2} + \frac{1}{s_{\nt}} \sum_{k=1}^{\nt} h_k (k-1). 
\end{IEEEeqnarray}

\subsection{Lower Bounds}
\label{sec:lower-bounds}

We first present the lower bounds.

\begin{proposition}[Lower Bound for $\alpha < \alpha_{\textnormal{th}}$]
  \label{prop:lower}
  If $\alpha < \alpha_{\textnormal{th}}$, the capacity is
  lower-bounded as
  \begin{IEEEeqnarray}{c}
    \C_{\trans{\vect{h}},\sigma^2}(\amp,\alpha\amp)
    \ge \frac{1}{2}\log\left( 1 + \frac{\amp^2 s_{\nt}^2}{2\pi
        e\sigma^2} \ope^{2\nu} \right)
    \label{eq:lower}
  \end{IEEEeqnarray}
  with
  \begin{IEEEeqnarray}{c}
    \nu \eqdef 	\sup_{\lambda\in\left(
        \max\{0,\frac{1}{2}+\alpha-\alpha_{\textnormal{th}}\}, 
        \min\{\frac{1}{2},\alpha\}\right) }  \Biggl\{
    1 - \log\frac{\mu(\lambda)}{1-\ope^{-\mu(\lambda)}}  
    - \frac{\mu(\lambda)\ope^{-\mu(\lambda)}}{1-\ope^{-\mu(\lambda)}}  
    - \relD\left(\vect{p} \middle\| \frac{\vect{h}}{s_{\nt}}
    \right)\Biggr\}, \nonumber\\*
    \label{eq:18}
  \end{IEEEeqnarray}
  where $\mu(\lambda)$ is the unique positive solution to the
  following equation in $\mu$:
  \begin{IEEEeqnarray}{c}
    \frac{1}{\mu} - \frac{\ope^{-\mu}}{1-\ope^{-\mu}}
    = \lambda;
    \label{eq:8}
  \end{IEEEeqnarray}
  and where
  \begin{IEEEeqnarray}{c}
    \subnumberinglabel{eq:pk}
    p_k = \frac{h_k a^k}{\sum_{j=1}^{\nt}
      h_{j} a^{j}}, \qquad k\in\{1,\ldots,\nt\}, 
  \end{IEEEeqnarray}
  with $a$ being the unique positive solution to 
  \begin{IEEEeqnarray}{c}
    \IEEEyessubnumber
    \frac{ \sum_{k=1}^{\nt}  h_k k a^k }{\sum_{j=1}^{\nt}
      h_j a^j} = \alpha-\lambda+1.
  \end{IEEEeqnarray}
\end{proposition}
\begin{IEEEproof}
  See Section~\ref{sec:finite-snr-lower-bound}.
\end{IEEEproof}
\bigskip

\begin{proposition}[Lower Bound for $\alpha \geq \alpha_{\textnormal{th}}$]
  \label{prop:lower2}
  If $\alpha \ge \alpha_{\textnormal{th}}$, the capacity is
  lower-bounded as
  \begin{IEEEeqnarray}{rCl}
    \C_{\trans{\vect{h}},\sigma^2}(\amp,
    \alpha\amp)   
    & \ge & \frac{1}{2} \log
    \left(1+\frac{s_{\nt}^2\amp^2}{2\pi e \sigma^2}\right).
    \label{eq:asympcaplower2}
  \end{IEEEeqnarray}
\end{proposition}
\begin{IEEEproof}
  See Section~\ref{sec:finite-snr-lower-bound2}.
\end{IEEEproof}
\bigskip

The lower bounds are obtained by choosing the inputs so as to maximize
the differential entropy $\hh(\Xs)$ under the constraints
\eqref{eq:newconstraints}. By Lemma~\ref{lem:energy_efficient}, this
means that for $\alpha < \alpha_{\textnormal{th}}$ we let with
probability $p_k$, $k\in\{1,\ldots,\nt\}$,
\begin{IEEEeqnarray}{l} 
  \subnumberinglabel{eq:expo}
  X_j=\amp, \qquad j\in\{1,\ldots,
  k-1\},  
  \label{eq:expoa}
  \\*
  X_k \sim \textnormal{truncated exponential of parameter } \mu(\lambda),
  \IEEEeqnarraynumspace 
  \label{eq:expob} 
  \\*
  X_j=0, \qquad j\in\{k+1,\ldots,\nt\},
  \label{eq:expoc} 
\end{IEEEeqnarray} 
where $\mu(\lambda)$ is the unique positive solution to \eqref{eq:8}
and $\vect{p}$ is given in \eqref{eq:pk}.  This
choice results in a concatenation of $\nt$ truncated exponentials for
$\Xs$.
	
For $\alpha > \alpha_{\textnormal{th}}$, we replace the truncated
exponential distribution in \eqref{eq:expob} by a uniform distribution
and choose the probability vector $\vect{p}$ as $p_k = h_k/s_{\nt}$, for
$k\in\{1,\ldots,\nt\}$. This choice yields a uniform distribution on
$[0,s_{\nt} \amp]$ for $\Xs$.
	
That the described choices maximize $\hh(\Xs)$ under constraints
\eqref{eq:newconstraints} can be proved, e.g., using
\cite[Thm.~12.1.1]{coverthomas06_1}.


\subsection{Upper Bounds}
\label{sec:upper-bounds}

We next present our upper bounds. First we present a simple upper
bound by the SISO capacity.

\begin{proposition}[Upper Bound by SISO Capacity]
  \label{prp:SISO}
  The capacity is upper-bounded as
  \begin{IEEEeqnarray}{rCl}
    \C_{\trans{\vect{h}},\sigma^2}(\amp,
    \alpha\amp)
    & \le & \C_{1,\sigma^2}\left( s_{\nt}\amp,\frac{s_{\nt}\amp}{2}
    \right). 
    \label{eq:asympcapupper2}
  \end{IEEEeqnarray}
\end{proposition}
\begin{IEEEproof}
  The bound follows by the equivalence in \eqref{eq:MISOpeakonly}, and
  because the capacity is nondecreasing in the parameter $\alpha$ (as
  the transmitter can always choose not to use all of its available
  power).
\end{IEEEproof}
\bigskip

Note that the SISO capacity
$\C_{1,\sigma^2}\left(s_{\nt}\amp,\frac{s_{\nt}\amp}{2}\right)$ is
itself unknown to date. Upper bounds on it were given in
\cite{lapidothmoserwigger09_7, smith71_1, mckellips04_1,
  thangarajkramerbocherer17_1}. In fact, under a peak-power constraint
$s_{\nt}\amp$, the average-power constraint $\frac{s_{\nt}\amp}{2}$ is
not active.

Our next upper bound, like Proposition~\ref{prp:SISO}, is valid for
all values of $\alpha < \frac{\nt}{2}$. It depends on the maximum
variance of $\Xs$ under constraints~\eqref{eq:newconstraints}:
\begin{IEEEeqnarray}{c}
  \Vmax (\amp,\alpha \amp) \eqdef \max_{Q_{\Xs}}
  \E{\bigl(\Xs-\eE{\Xs} \bigr)^2}, 
  \label{eq:max1}
\end{IEEEeqnarray}
where the maximization is over all distributions on $\Xs\geq 0$
that satisfy constraints \eqref{eq:newconstraints}. This maximum
variance is easily calculated numerically using the following
lemma. 
	
\begin{lemma}
  \label{lem:maximum_variance}
  Consider the maximum variance $\Vmax(\amp, \alpha \amp)$ as defined
  in \eqref{eq:max1}.
  \begin{enumerate}
  \item The maximum variance can be achieved by restricting $Q_{\Xs}$
    to the support set
    \begin{IEEEeqnarray}{c}
      \{0, s_1\amp , s_2\amp, \ldots, s_{\nt}\amp \}.\label{eq:nt+1}
    \end{IEEEeqnarray}
    
  \item The maximum variance satisfies
    \begin{IEEEeqnarray}{c}
      \Vmax(\amp,\alpha \amp) = \amp^2 \gamma
    \end{IEEEeqnarray}
    where 
    \begin{IEEEeqnarray}{c}
      \label{eq:delta}
      \gamma\eqdef \max_{\substack{{q_1,\ldots,\ q_{\nt} \geq 0}\colon
          \\[0.5ex] \sum_{k=1}^{\nt}  q_k \leq 1 \\[0.5ex]
          \sum_{k=1}^{\nt} k \cdot q_k  \leq \alpha}}
      \left(\sum_{k=1}^{\nt}  s_k^2 q_k - \bigg(\sum_{k=1}^{\nt}  s_k
        q_k\bigg)^2\right). 
    \end{IEEEeqnarray}
  \end{enumerate}
\end{lemma}
\begin{IEEEproof}
  See Appendix~\ref{app:variance}.
\end{IEEEproof}
\bigskip

\begin{table}[tb]
  \centering
  \caption{Maximum variance for different channel coefficients}
  \label{tab:vmax}
  \vspace{3mm}
  \begin{IEEEeqnarraybox}[\mystrut]{l"l"l"l}
    \hline\hline
    \IEEEeqnarraymulticol{1}{t}{channel gains}
    & \IEEEeqnarraymulticol{1}{c}{\alpha}
    & \IEEEeqnarraymulticol{1}{c}{\Vmax}
    & \IEEEeqnarraymulticol{1}{t}{$Q_{\Xs}$ achieving $\Vmax$}
    \\\hline
    \vect{h}=(3,2.2,0.1) & 0.9 & 6.6924 \amp^2
    & Q_{\Xs}(0)=0.55,\; Q_{\Xs}(s_2\amp)=0.45
    \\
    \vect{h}=(3,2.2,1.1) & 0.7 & 7.1001\amp^2
    & Q_{\Xs}(0)=0.7667,\; Q_{\Xs}(s_3\amp)=0.2333
    \\
    \vect{h}=(3,1.5,0.3) & 0.95 & 5.1158\amp^2
    & Q_{\Xs}(0)=0.5907,\; Q_{\Xs}(s_2 \amp) =0.2780, \\
    &&& Q_{\Xs}(s_3 \amp)=0.1313
    \\\hline\hline
  \end{IEEEeqnarraybox}
\end{table}

For a three-LED MISO channel, some examples of variances $\Vmax$
are shown in Table~\ref{tab:vmax}. The last column of the table
indicates the probability mass function that achieves $\Vmax$.

\begin{proposition}[Upper Bound in Terms of $\Vmax$]
  \label{prop:simpler_upper}
  The capacity is upper-bounded as
  \begin{IEEEeqnarray}{rCl}
    \C_{\trans{\vect{h}},\sigma^2}(\amp,\alpha\amp)
    & \le & \frac{1}{2}\log \left( 1+ \frac{\Vmax(\amp,\alpha
        \amp)}{\sigma^2}\right).
    \label{eq:ub1}
  \end{IEEEeqnarray}
\end{proposition}
\begin{IEEEproof}
  Since $\Xs$ and $Z$ are independent, we know that the variance of
  $Y$ cannot exceed $\Vmax(\amp,\alpha\amp)+\sigma^2$, and therefore
  \begin{IEEEeqnarray}{c}
    \hh(Y) \le \frac{1}{2}\log 2\pi
    e\left(\Vmax(\amp,\alpha\amp)+\sigma^2\right). 
  \end{IEEEeqnarray}
  The bound follows by subtracting
  $\hh(Z) = \frac{1}{2} \log2\pi e \sigma^2$ from the above.
\end{IEEEproof}
\bigskip

Our last upper bound is more involved than the previous two. It holds
only when $\alpha<\alpha_\textnormal{th}$.

\begin{proposition}[Upper Bound for $\alpha < \alpha_{\textnormal{th}}$]
  \label{prop:finiteSNR}
  If $\alpha < \alpha_{\textnormal{th}}$, then capacity is
  upper-bounded as
  \begin{IEEEeqnarray}{rCl}
    \IEEEeqnarraymulticol{3}{l}{%
      \C_{\trans{\vect{h}},\sigma^2}(\amp,\alpha\amp)
    }\nonumber\\*\;\;%
    & \le & \sup_{\vect{p}} \inf_{\delta,\mu>0}\Biggl\{
    \frac{1}{2} \log \frac{\amp^2 s_{\nt}^2}{2\pi e \sigma^2}
    -\log \mu - \log \left(1-2\Qf{\frac{\delta}{\sigma}}\right)
    + \Qf{\frac{\delta}{\sigma}}
    + \frac{\delta}{\sqrt{2\pi}\sigma} \ope^{-\frac{\delta^2}{2\sigma^2}} 
    \nonumber\\
    && \qquad\qquad\;\; - \relD\left(\vect{p} \middle\|
      \frac{\vect{h}}{s_{\nt}}\right) 
    + \sum_{k=1}^{\nt} p_k \log \left( \ope^{\frac{\mu \delta}{ 
          \amp h_k}} - \ope^{-\mu\left(1+\frac{\delta}{\amp
            h_k}\right)}\right) 
    \nonumber\\
    && \qquad\qquad\;\; +\> \frac{\mu\sigma}{\amp\sqrt{2\pi}}
    \sum_{k=1}^{\nt} 
    \frac{p_k}{h_k} \left(\ope^{-\frac{\delta^2}{2\sigma^2}}
      -\ope^{-\frac{\left(\amp h_k+\delta\right)^2}{2\sigma^2}}\right)
    + \mu \left( \alpha - \sum_{k=1}^{\nt} p_k (k-1) \right)  
    \Biggr\},
    \label{eq:ub2}
    \IEEEeqnarraynumspace
  \end{IEEEeqnarray}
  where the supremum is over all probability vectors
  $\vect{p}=(p_1,\ldots, p_{\nt})$ satisfying
  \begin{IEEEeqnarray}{c}
    \label{eq:m22}
    \sum_{k=1}^{\nt} p_k (k-1) \leq \alpha.
  \end{IEEEeqnarray}
\end{proposition} 
\begin{IEEEproof}
  See Section~\ref{sec:finiteSNR}.
\end{IEEEproof}
\bigskip

\begin{figure}[htbp]
  \centering
  \begin{tikzpicture}[xscale=0.33,yscale=2]
    \draw [dotted,xstep=5,ystep=0.5] (-15,-0.1) grid (20,5);
    \draw [thick] (-15,-0.1) rectangle (20,5);
    \node[rotate=90] at (-18,2.5) {$\C_{\trans{\vect{h}},\sigma^2}(\amp,\alpha\amp)$ [nats]}; 
    \node at (2.5,-0.6) {$\amp$ [dB]};
    \foreach \i in {-15,-10,-5,0,5,10,15,20}{
      \node [below] at (\i,-0.1) {$\i$};
    };
    \foreach \i in {0,0.5,...,5}{
      \node [left] at (-15,\i) {$\i$};
    };
    \draw [thick,black] plot file{num0.mat};
    \draw [thick,magenta,densely dashdotted] plot file{num1.mat};
    \draw [thick,blue,densely dashed] plot file{num3.mat};
    \draw [thick,red,solid] plot file{num4.mat};
    \draw [thick,cyan,densely dotted] plot file{num10.mat};
    \draw [thick,violet,densely dashdotdotted] plot file{num11.mat};

    \filldraw [fill=white,draw=black] (-13.75,2.75) rectangle (3.25,4.75);
    \draw [thick,black] (-13.25,3)--(-11.25,3);
    \node [right] at (-11.05,3) {\footnotesize
      Lower Bound (Prop.~\ref{prop:lower})}; 
    \draw [thick,red,solid] (-13.25,3.3)--(-11.25,3.3);
    \node [right] at (-11.05,3.3) {\footnotesize
      Upper Bound (Prop.~\ref{prop:finiteSNR})}; 
    \draw [thick,violet,densely dashdotdotted] (-13.25,3.6)--(-11.25,3.6);
    \node [right] at (-11.05,3.6) {\footnotesize
      SISO-U.B. (Prop.~\ref{prp:SISO}) with
      \cite[(17)]{thangarajkramerbocherer17_1}};
    \draw [thick,cyan,densely dotted] (-13.25,3.9)--(-11.25,3.9);
    \node [right] at (-11.05,3.9) {\footnotesize
      SISO-U.B. (Prop.~\ref{prp:SISO}) with
      \cite{mckellips04_1}};
    \draw [thick,blue,densely dashed] (-13.25,4.2)--(-11.25,4.2);
    \node [right] at (-11.05,4.2) {\footnotesize
      SISO-U.B. (Prop.~\ref{prp:SISO}) with
      \cite[(20)]{lapidothmoserwigger09_7}};
    \draw [thick,magenta,densely dashdotted] (-13.25,4.5)--(-11.25,4.5);
    \node [right] at (-11.05,4.5) {\footnotesize
      $\Vmax$-U.B. (Prop.~\ref{prop:simpler_upper})}; 
  \end{tikzpicture}
  \caption{Bounds on the capacity of the MISO channel with three LEDs
    and channel gains $\vect{h}=(3,2,1.5)$ for the case
    $\alpha=0.6$. Note that the threshold of this channel is
    $\alpha_{\textnormal{th}}=1.2692$ (and $\frac{\nt}{2}=1.5$). The
    SISO upper bound from Proposition~\ref{prp:SISO} is plotted in
    combination with three known SISO capacity bounds taken from
    \cite{mckellips04_1, lapidothmoserwigger09_7,
      thangarajkramerbocherer17_1}. Note that the bound from
    \cite{thangarajkramerbocherer17_1} is only valid for
    $\amp \le -1.97$ dB.}
  \label{fig:numerical}
\end{figure}
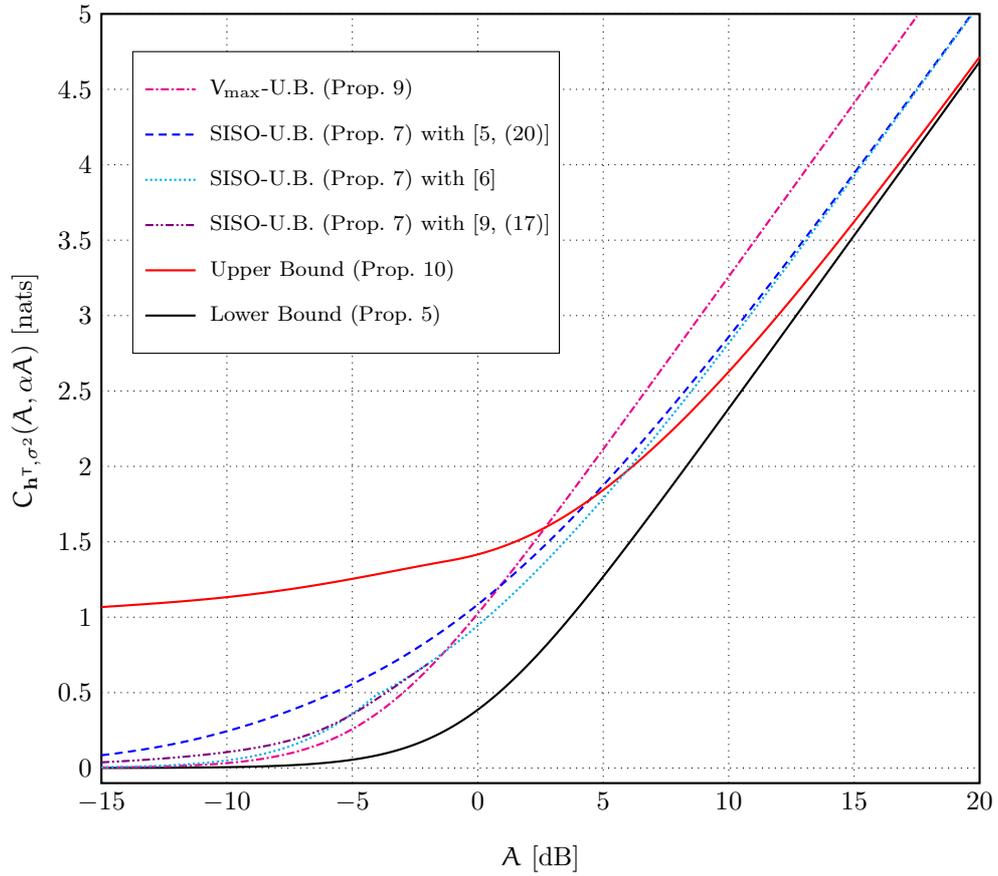

Figure~\ref{fig:numerical} shows our lower and upper
bounds for the MISO channel with gains $\vect{h}=( 3, 2, 1.5)$.
It suggests that the upper bound in Proposition~\ref{prop:finiteSNR}
and the lower bound in Proposition~\ref{prop:lower} are asymptotically
tight as $\amp \to \infty$. This is indeed the case, as we show in
Proposition~\ref{thm:main} below. In the low-SNR regime, all our upper
and lower bounds, except the upper bound in
Proposition~\ref{prop:finiteSNR}, tend to zero, but not with the same
slope. To characterize the capacity slope at low SNR, we shall use an
asymptotic lower bound from \cite{prelovvandermeulen93_1}; see
Proposition~\ref{thm:low_SNR}.

\subsection{Asymptotic Results}
\label{sec:asymptotic-results}

\begin{proposition}[High-SNR Asymptotics]
  \label{thm:main}
  If $\alpha \ge \alpha_{\textnormal{th}}$, then
  \begin{IEEEeqnarray}{c}
    \lim_{\amp\to\infty} \bigl\{ \C_{\trans{\vect{h}},\sigma^2}(\amp,
    \alpha\amp) - \log \amp \bigr\} 
    = \frac{1}{2} \log \frac{s_{\nt}^2}{2\pi e\sigma^2}.
    \label{eq:12}
  \end{IEEEeqnarray}
  If $\alpha<\alpha_{\textnormal{th}}$, then
  \begin{IEEEeqnarray}{rCl}
    \IEEEeqnarraymulticol{3}{l}{%
      \lim_{\amp\to\infty} \bigl\{ \C_{\trans{\vect{h}},\sigma^2}(\amp,
      \alpha\amp) - \log \amp \bigr\} 
    }\nonumber\\*\quad%
    & = & \frac{1}{2}\log \frac{s_{\nt}^2}{2\pi e \sigma^2}
    \nonumber\\
    && + \sup_{\lambda\in\left(
        \max\{0,\frac{1}{2}+\alpha-\alpha_{\textnormal{th}}\}, 
        \min\{\frac{1}{2},\alpha\}\right)}  \left\{
      1 - \log\frac{\mu(\lambda)}{1-\ope^{-\mu(\lambda)}}
      - \frac{\mu(\lambda)\ope^{-\mu(\lambda)}}{1-\ope^{-\mu(\lambda)}}  
      - \relD\left(\vect{p} \middle\| \frac{\vect{h}}{s_{\nt}}
      \right)\right\},
    \nonumber\\*
    \label{eq:asympcap}
  \end{IEEEeqnarray}
  where $\mu(\lambda)$ and $\vect{p}=(p_1,\ldots, p_{\nt})$ are the
  same as in Proposition~\ref{prop:lower}.
\end{proposition}
\begin{IEEEproof}
  Achievability of \eqref{eq:12} and \eqref{eq:asympcap} follows
  immediately from the lower bounds in Propositions~\ref{prop:lower2}
  and \ref{prop:lower}, respectively. The converse to \eqref{eq:12} is
  based on the upper bound \eqref{eq:asympcapupper2} in
  Proposition~\ref{prp:SISO} and the high-SNR analysis of the SISO
  capacity in \cite[Cor.~6]{lapidothmoserwigger09_7}. The converse to
  \eqref{eq:asympcap} is based on Proposition~\ref{prop:finiteSNR} and
  is given in detail in Section~\ref{sec:asymptotic}.
\end{IEEEproof}
\bigskip

\begin{figure}[htbp]
  \centering
  \begin{tikzpicture}[xscale=7,yscale=1.5]
    \draw [thick] (0,0) rectangle (1,4);
    \draw [dotted,xstep=0.1,ystep=0.5] (0,0) grid (1,4);
    \node[rotate=90] at (-0.13,2) {Gap to
      $\C_{\trans{\vect{h}},\sigma^2}(\amp,\amp)$ [nats]}; 
    \node at (0.5,-0.5) {$\alpha$};
    \foreach \i in {0,0.1,0.2,0.3,0.4,0.5,0.6,0.7,0.8,0.9,1}{
      \node [below] at (\i,0) {$\i$};
    };
    \foreach \i in {0,0.5,1,...,4}{
      \node [left] at (0,\i) {$\i$};
    };
    \draw [thick] plot file{alpha_gap.mat};
  \end{tikzpicture}
  \caption{The supremum in \eqref{eq:asympcap} as a function of
    $\alpha$, for a MISO channel with two LEDs and parameters $h_1=3$
    and $h_2=1$. This supremum is the asymptotic capacity gap to the
    case with no average-power constraint
    $\C_{\trans{\vect{h}},\sigma^2}(\amp,\amp)$.}
  \label{fig:1}
\end{figure}
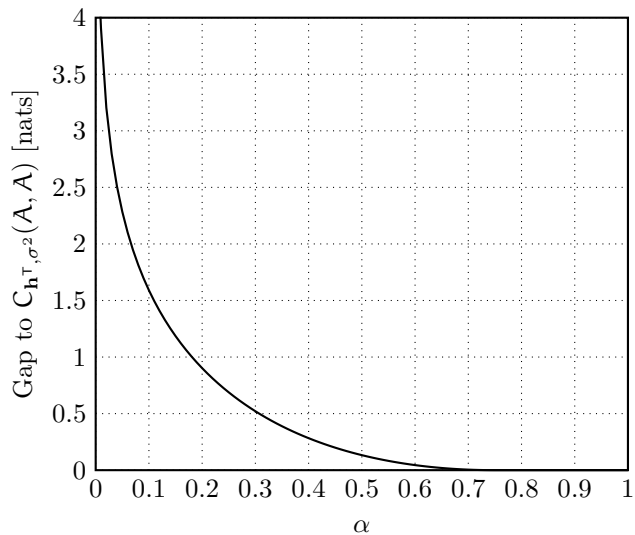

\begin{example*}
  Consider a MISO channel with two LEDs and with channel parameters
  $h_1=3$ and $h_2=1$. We plot the supremum on the RHS of
  \eqref{eq:asympcap} against $\alpha$ in Figure~\ref{fig:1}. Note
  that this supremum characterizes the capacity gap to the case with
  no average-power constraint in the high-SNR limit. As expected, the
  gap becomes zero when $\alpha$ reaches
  $\alpha_{\textnormal{th}}=0.75$, and approaches infinity when
  $\alpha$ tends to zero.
\end{example*}

\bigskip

We now turn to the low-SNR asymptotic regime.  In this regime the
capacity is determined by $\Vmax(\amp, \alpha \amp)$.
\begin{proposition}[Low-SNR Asymptotics]
  \label{thm:low_SNR}
  The low-SNR asymptotic capacity is
  \begin{IEEEeqnarray}{c}
    \lim_{\amp\downarrow 0} \frac{
      \C_{\trans{\vect{h}},\sigma^2}(\amp, 
      \alpha\amp)}{ \amp^2/\sigma^2 }  
    = \frac{\gamma}{2},
    \label{eq:21m}
  \end{IEEEeqnarray}
  where $\gamma$ is defined in \eqref{eq:delta}.
\end{proposition}
\begin{IEEEproof}
  The converse follows immediately from the upper bound \eqref{eq:ub1}
  in Proposition~\ref{prop:simpler_upper}.  Achievability follows from
  \cite[Thm.~2]{prelovvandermeulen93_1}, which states that
  \begin{IEEEeqnarray}{c}
    \label{eq:thm2cond}
    \C_{\trans{\vect{h}},\sigma^2}(\amp,
    \alpha\amp) \geq  \frac{\Vmax(\amp, \alpha \amp)}{2\sigma^2} +
    o(\amp^2), 
  \end{IEEEeqnarray} 
  where $o(\amp^2)$ decreases to 0 faster than $\amp^2$, i.e., 
  \begin{IEEEeqnarray}{c}
    \lim_{\amp\downarrow 0} \frac{o(\amp^2)}{\amp^2}=0. 
  \end{IEEEeqnarray}
  Note that the MISO channel under consideration in this paper
  satisfies the technical conditions A--F in
  \cite{prelovvandermeulen93_1}.
\end{IEEEproof}
\bigskip

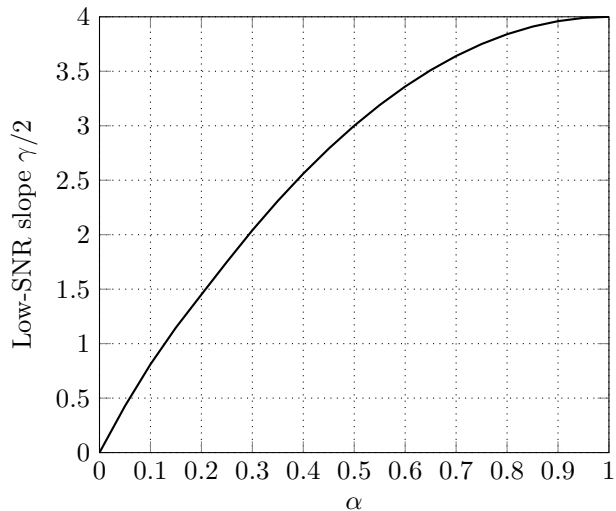
\begin{figure}[htbp]
  \centering
   \begin{tikzpicture} [every pin/.style={fill=white},scale=1]
    \begin{axis}[scale=1,
      width=0.5\columnwidth,
      scale only axis,
      xmin=0,
      xmax=1,
      xtick={0,0.1,...,1.1},
      xmajorgrids,
      xlabel={$\alpha$},
      ymin=0,
      ymax=4,
      ytick={0,0.5,...,4},
      ymajorgrids,
      ylabel={Low-SNR slope $\gamma/2$},
      grid style={dotted,black}
      ]
      
      \addplot[color=black,solid,thick]
      table[row sep=crcr]{
	0.00000    0\\
	0.05000    0.42750\\
	0.10000     0.81000 \\
	0.15000    1.14750 \\
	0.20000     1.45000 \\
	0.25000      1.75000 \\
	0.30000    2.04000\\
	0.35000    2.31000\\
	0.40000    2.56000 \\
	0.45000    2.79000\\
	0.50000    3.00000\\
	0.55000    3.19000 \\
	0.60000    3.36000\\
	0.65000     3.51000\\
	0.70000    3.64000 \\
	0.75000    3.75000  \\
	0.80000    3.84000\\
	0.85000       3.91000 \\
	0.90000    3.96000\\
	0.95000    3.99000\\
	1.0000    4\\
      };
    \end{axis}
  \end{tikzpicture}
  \caption{The low-SNR asymptotic slope $\gamma/2$, see
    \eqref{eq:21m}, is depicted as a function of $\alpha \in (0,1)$
    for the two-LED MISO channel with gains $h_1=3$ and $h_2=1$ (the
    same channel as considered in Figure~\ref{fig:1}).}
  \label{fig:2}
\end{figure}

\begin{example*}
  We return to the example from before and reconsider the two-LED
  MISO channel of Figure~\ref{fig:1} with channel gains $h_1=3$ and
  $h_2=1$.  Figure~\ref{fig:2} plots the low-SNR slope of its capacity
  $\gamma/2$ as a function of the parameter $\alpha$.  We notice that
  the low-SNR slope $\gamma/2$ does not reach a constant value when
  $\alpha \geq\alpha_{\textnormal{th}}$, but it is strictly increasing
  for all values of $\alpha < \frac{\nt}{2}$.
\end{example*}

\section{Achievability Proofs (Proofs of  Lower Bounds)}
\label{sec:lower-bound}

\subsection{Proof of Proposition~\ref{prop:lower}}
\label{sec:finite-snr-lower-bound}

Fix
\begin{IEEEeqnarray}{c}
  \lambda\in\left(
    \max\left\{0,\frac{1}{2}+\alpha-\alpha_{\textnormal{th}}\right\},
    \min\left\{\frac{1}{2},\alpha\right\}\right)
\end{IEEEeqnarray}
and choose a probability vector $\vect{p}=(p_1,\ldots, p_{\nt})$
satisfying
\begin{IEEEeqnarray}{c}
  \label{eq:choice_p}
  \sum_{k=1}^{\nt} p_k (k-1) = \alpha - \lambda.
\end{IEEEeqnarray} 
Such a choice always exists. To see this, first note
$0<\alpha-\lambda<\nt/2$. When we choose $\vect{p}=(1,0,\ldots,0)$,
the LHS of \eqref{eq:choice_p} equals 0; when we choose
$\vect{p}=(0,\ldots,0,1)$, it equals $\nt-1$, which is larger than or
equal to $\nt/2$ for all $\nt\ge 2$. The existence of a $\vect{p}$
satisfying \eqref{eq:choice_p} then follows by the continuity of the
LHS of \eqref{eq:choice_p} in~$\vect{p}$.

Now let $\Xs$ have the following probability density function (PDF):
for every $k\in\{1,\ldots, \nt\}$, for
$\bar{x} \in (s_{k-1}\amp, s_k \amp]$,
\begin{IEEEeqnarray}{c}
  f_{\Xs}(\bar{x}) = \frac{p_k}{h_k\amp} \,
  \frac{\mu(\lambda)}{1-\ope^{-\mu(\lambda)}} 
  \ope^{-\frac{\mu(\lambda) (\bar{x}-s_{k-1}\amp)}{h_{k}\amp}},
  \label{eq:PDF}
\end{IEEEeqnarray}
where $\mu(\lambda)$ is the unique positive solution to \eqref{eq:8};
and let $U$ be the random variable (RV) defined in \eqref{eq:U}
corresponding to this choice of $\Xs$. It is easy to check that the
choice \eqref{eq:PDF} indeed constitutes a PDF. Further, it satisfies
\eqref{eq:newconstraints} because it has positive probability only on
the interval $[0, s_{\nt}\amp]$, and because
\begin{IEEEeqnarray}{rCl}
  \IEEEeqnarraymulticol{3}{l}{%
    \sum_{k=1}^{\nt} p_{k}\left(\frac{\E{ \Xs|U=k} - \amp
        s_{k-1}}{h_{k}} +(k-1)\amp \right) 
  }\nonumber\\*\quad%
  & = & \sum_{k=1}^{\nt} p_k \left( \left(\frac{1}{\mu(\lambda)} -
      \frac{\ope^{-\mu(\lambda)}}{1-\ope^{-\mu(\lambda)}}
    \right)\amp+(k-1)\amp \right) 
  \\
  & = & \sum_{k=1}^{\nt} p_k \left(\lambda\amp+(k-1)\amp
  \right)
  \label{eq:m40}
  \\
  & = & \lambda \amp + \sum_{k=1}^{\nt} p_k(k-1)\amp
  \\
  & = & \alpha\amp.
  \label{eq:m41}
\end{IEEEeqnarray}
Here, \eqref{eq:m40} follows from \eqref{eq:8}, and \eqref{eq:m41}
from \eqref{eq:choice_p}. 
We then evaluate the mutual information in~\eqref{eq:C_equiv} for this
$\Xs$ and use the Entropy Power Inequality (EPI)
\cite[Thm.~17.7.3]{coverthomas06_1} to obtain
\begin{IEEEeqnarray}{rCl}
  \C_{\trans{\vect{h}},\sigma^2}(\amp,\alpha\amp)
  & \ge & \II(\Xs; \Xs + Z)
  \label{eq:m1}
  \\
  & = & \hh(\Xs + Z) - \hh(Z)
  \\
  & \ge & \frac{1}{2}\log\bigl( \ope^{2\hh(\Xs)}
  + \ope^{2\hh(Z)} \bigr) - \hh(Z)
  \label{eq:9}
  \\
  & = & \frac{1}{2}\log\bigl( \ope^{2\hh(\Xs)}
  + 2\pi e\sigma^2 \bigr) - \frac{1}{2}\log 2\pi e\sigma^2
  \\
  & = & \frac{1}{2}\log\left(1 + \frac{\ope^{2\hh(\Xs)}}{2\pi
      e\sigma^2} \right).
  \label{eq:13}
\end{IEEEeqnarray}
We next evaluate the differential entropy for the chosen $\Xs$:
\begin{IEEEeqnarray}{rCl}
  \hh(\Xs)
  & = & \HH(U) - \underbrace{\HH(U|\Xs)}_{=\,0}{} + \hh(\Xs|U)
  \\ 
  & = & \HH(\vect{p}) + \sum_{k=1}^{\nt} p_k \hh(\Xs|U=k)
  \label{eq:19b} 
  \\
  & = & \HH(\vect{p}) + \sum_{k=1}^{\nt} p_k \log h_k +\log\amp 
  - \log\frac{\mu(\lambda)}{1-\ope^{-\mu(\lambda)}}  
  + 1 - \frac{\mu(\lambda)\ope^{-\mu(\lambda)}}{1-\ope^{-\mu(\lambda)}}
  \label{eq:15}
  \\
  & = & - \relD\left(\vect{p} \middle\| \frac{\vect{h}}{s_{\nt}}\right)
  + \log s_{\nt}\amp - \log\frac{\mu(\lambda)}{1-\ope^{-\mu(\lambda)}}  
  + 1 - \frac{\mu(\lambda)\ope^{-\mu(\lambda)}}{1-\ope^{-\mu(\lambda)}},
  \label{eq:7}
\end{IEEEeqnarray}  
where \eqref{eq:15} follows from the differential entropy expression
of a truncated exponential RV.
 
The proposition then follows by plugging \eqref{eq:7} into
\eqref{eq:13} and by maximizing the lower bound over the choice of the
probability vector $\vect{p}=(p_1,\ldots, p_{\nt})$ subject to
constraint~\eqref{eq:choice_p} and then maximizing over the choice of
$\lambda$.

It only remains to show that the optimal choice of $\vect{p}$ in
\eqref{eq:18} indeed has the form given in \eqref{eq:pk}. To that goal
note that the first three terms inside the supremum in \eqref{eq:18}:
\begin{IEEEeqnarray}{c}
  1 - \log\frac{\mu(\lambda)}{1-\ope^{-\mu(\lambda)}}  
  - \frac{\mu(\lambda)\ope^{-\mu(\lambda)}}{1-\ope^{-\mu(\lambda)}}
\end{IEEEeqnarray}
correspond to the differential entropy of a truncated exponential RV
$V\in[0,1]$ with mean $\E{V}=\lambda$ for
$\lambda\in\left(0,\frac{1}{2}\right)$. This expression is
monotonically strictly increasing in $\lambda$ and approaches its
maximum for $\lambda \to\frac{1}{2}$ (in which case $V$ approaches a
uniformly distributed random variable on $[0,1]$).

The last term inside the supremum in
\eqref{eq:18}:
\begin{IEEEeqnarray}{c}
  - \relD\left(\vect{p} \middle\| \frac{\vect{h}}{s_{\nt}}\right)
  \label{eq:19}
\end{IEEEeqnarray}
is nonpositive and equals zero if, and only if, $\vect{p} =
\frac{\vect{h}}{s_{\nt}}$, in which case
\begin{IEEEeqnarray}{c}
  \sum_{k=1}^{\nt} p_k (k-1) = \alpha_{\textnormal{th}} - \frac{1}{2}.    
\end{IEEEeqnarray}
Thus, as long as
\begin{IEEEeqnarray}{c}
  \alpha - \lambda < \alpha_{\textnormal{th}} - \frac{1}{2}
\end{IEEEeqnarray}
the constraint~\eqref{eq:choice_p} is active and
\cite[Probl.~12.2]{coverthomas06_1} tells us that the unique optimal
$\vect{p}$ that maximizes \eqref{eq:19} and simultaneously satisfies
\eqref{eq:choice_p} has the form given in \eqref{eq:pk}.

\subsection{Proof of Proposition~\ref{prop:lower2}}
\label{sec:finite-snr-lower-bound2}

We choose $\Xs$ to be uniform on $[0,s_{\nt} \amp]$. Let $U$ be the RV
defined in \eqref{eq:U} for this choice of $\Xs$, then $\Prv{U=k}=p_k$
where
\begin{IEEEeqnarray}{c}
  p_k \eqdef \frac{h_k}{s_{\nt}}, \quad k=1, \ldots, \nt,
\end{IEEEeqnarray}
and, conditional on $U=k$, $\Xs$ is uniform over
$(s_{k-1}\amp,s_k\amp]$.  The chosen
$\Xs$ satisfies \eqref{eq:newconstraints} because it has positive
probability only on the interval $[0, s_{\nt} \amp]$ and because
\begin{IEEEeqnarray}{rCl}
  \sum_{k=1}^{\nt} p_{k}\left(\frac{\E{ \Xs|U=k} - \amp
      s_{k-1}}{h_{k}} +(k-1)\amp \right) 
  & = & \sum_{k=1}^{\nt} p_k \left( \frac{\amp}{2} + (k-1)\amp \right)
  \\
  & = & \frac{\amp}{2} + \amp\sum_{k=1}^{\nt} \frac{h_k}{s_{\nt}} (k-1) 
  \\
  & = &  \left(\frac{1}{2} + \frac{1}{s_{\nt}}\sum_{k=1}^{\nt}h_k
    (k-1)\right)\amp  
  \IEEEeqnarraynumspace
  \\
  & = & \alpha_{\textnormal{th}} \amp
  \label{eq:3}
  \\
  & \le & \alpha\amp,
  \label{eq:6}
\end{IEEEeqnarray}
where \eqref{eq:3} follows from the definition of
$\alpha_{\textnormal{th}}$ in \eqref{eq:alpha48}. Like in
\eqref{eq:m1}--\eqref{eq:13}, we obtain
\begin{IEEEeqnarray}{c}
\C_{\trans{\vect{h}},\sigma^2}(\amp,\alpha\amp)
  \ge \frac{1}{2}\log\left(1 + \frac{\ope^{2\hh(\Xs)}}{2\pi
      e\sigma^2} \right)\label{eq:ligong13}
\end{IEEEeqnarray}
for the above choice of $\Xs$. Since $\Xs$ is uniform, we have
\begin{IEEEeqnarray}{rCl}
  \hh(\Xs)
  & = & \log s_{\nt} \amp.
\end{IEEEeqnarray}
Plugging this into \eqref{eq:ligong13} yields the desired bound.

\section{Proof of Proposition~\ref{prop:finiteSNR}}
\label{sec:finiteSNR}

Choose $\Xs\sim Q_{\Xs}^*$ to be a maximizer of the capacity
expression in \eqref{eq:C_equiv}.  That means in particular that
$Q_{\Xs}^*$ satisfies \eqref{eq:newconstraints}, and thus that the
conditional expectations
\begin{IEEEeqnarray}{c}
  \alpha_k^* \eqdef
  \Econd[Q^*_{\Xs}]{\frac{\Xs-s_{k-1}\amp}{h_k\amp}}{U=k}  
  \label{eq:16}
\end{IEEEeqnarray}
satisfy
\begin{IEEEeqnarray}{c}
  \label{eq:17}
  \sum_{k=1}^{\nt} p_{k}^*\bigl(\alpha_{k}^*+(k-1) \bigr) \le \alpha,
\end{IEEEeqnarray}
where $U$ is the RV defined in \eqref{eq:U} and $p_k^*=\Prv{U=k}$.
   
The capacity is upper-bounded as follows:
\begin{IEEEeqnarray}{rCl}
  \C_{\trans{\vect{h}},\sigma^2}(\amp,\alpha\amp)
  & = & \II(\Xs; \Xs+Z)
  \\
  & \le & \II(\Xs; \Xs+Z,U)
  \\
  & = & \II(\Xs; U) + \II(\Xs; \Xs+Z | U)
  \\
  & = & \HH(U) - \underbrace{\HH(U|\Xs)}_{=\,0}{}
  + \sum_{k=1}^{\nt} p_k^* \II(\Xs; \Xs+Z | U=k) 
  \\
  & = & \HH(\vect{p}^*)  + \sum_{k=1}^{\nt} p_k^* \II(\Xs; \Xs+Z | U=k)
  \\
  & = & \HH(\vect{p}^*)  + \sum_{k=1}^{\nt} p_k^* \II(X_k; h_k X_k+Z |
  U=k) 
  \\
  & \le & \HH(\vect{p}^*) + \sum_{k=1}^{\nt} p_k^*
  \C_{1,\sigma^2}(h_k\amp,\alpha_k^* h_k \amp),
  \label{eq:m50}
\end{IEEEeqnarray}
where the inequality in \eqref{eq:m50} holds because, by
Lemma~\ref{lem:energy_efficient}, given $U=k$ we have
$\Xs=s_{k-1}\amp +h_k X_k$, where $X_k$ lies on the interval
$[0,\amp]$ and is of average power $\alpha_k^* \amp$.
  	
The SISO capacity $\C_{1,\sigma^2}( h_k \amp_k,\alpha_k^* h_k \amp)$
has been upper-bounded in \cite[Eq.~(12)]{lapidothmoserwigger09_7}.
Plugging this bound into \eqref{eq:m50} and performing some simple
bounding steps prove that for every choice of positive parameters
$\delta$ and $\mu$:
\begin{IEEEeqnarray}{rCl}
  \IEEEeqnarraymulticol{3}{l}{%
    \C_{\trans{\vect{h}},\sigma^2}(\amp,\alpha\amp)
  }\nonumber\\*\;\;%
  & \le & \HH(\vect{p}^*)
  + \sum_{k=1}^{\nt} p_k^* \log \left( \frac{\amp h_k}{\sigma}\cdot
    \frac{\ope^{\frac{\mu \delta}{ 
          \amp h_k}} - \ope^{-\mu\left(1+\frac{\delta}{\amp h_k}\right)}}
    {\sqrt{2\pi}\mu\left(1-2\Qf{\frac{\delta}{\sigma}}\right)}
  \right)
  - \frac{1}{2} + \Qf{\frac{\delta}{\sigma}}
  \nonumber\\
  && +\> \frac{\delta}{\sqrt{2\pi}\sigma}
  \ope^{-\frac{\delta^2}{2\sigma^2}}  
  + \frac{\mu\sigma}{\amp\sqrt{2\pi}} \sum_{k=1}^{\nt}
  \frac{p_k^*}{h_k} \left(\ope^{-\frac{\delta^2}{2\sigma^2}}
    -\ope^{-\frac{\left(\amp h_k+\delta\right)^2}{2\sigma^2}}\right)
  +  \mu \sum_{k=1}^{\nt} p_k^* \alpha_k^*
  \IEEEeqnarraynumspace
  \\
  & \le & 
  \sum_{k=1}^{\nt} p_k^*\log \frac{h_k}{p_k^*} + \log\amp
  - \frac{1}{2}\log (2\pi e  \sigma^2) - \log\mu
  - \log \left(1-2\Qf{\frac{\delta}{\sigma}}\right)
  \nonumber\\
  && + \sum_{k=1}^{\nt} p_k^* \log \left( \ope^{\frac{\mu \delta}{ 
        \amp h_k}} - \ope^{-\mu\left(1+\frac{\delta}{\amp
          h_k}\right)}\right) 
  + \Qf{\frac{\delta}{\sigma}}
  + \frac{\delta}{\sqrt{2\pi}\sigma} \ope^{-\frac{\delta^2}{2\sigma^2}} 
  \nonumber\\
  && +\> \frac{\mu\sigma}{\amp\sqrt{2\pi}} \sum_{k=1}^{\nt}
  \frac{p_k^*}{h_k} \left(\ope^{-\frac{\delta^2}{2\sigma^2}}
    -\ope^{-\frac{\left(\amp h_k+\delta\right)^2}{2\sigma^2}}\right)
  +  \mu \left( \alpha - \sum_{k=1}^{\nt} p_k^* (k-1) \right),
  \label{eq:23}
\end{IEEEeqnarray}
where \eqref{eq:23} follows from \eqref{eq:17} and by rearranging
terms. Since \eqref{eq:23} holds for all $\delta,\mu>0$, it must also
hold when we take the infimum of its RHS over $\delta,\mu>0$. Then we
relax this bound by further taking a supremum\footnote{Here we also
  need to make sure that $\vect{p}$ is chosen such that
  $\alpha - \sum_{k=1}^{\nt} p_k^* (k-1) \ge 0$.} over $\vect{p}$,
which establishes \eqref{eq:ub2}.

\section{Asymptotic High-SNR Analysis---Converse Proof to \texorpdfstring{\eqref{eq:asympcap}}{(48)}}
\label{sec:asymptotic}

Recall that, throughout this section, we are only concerned with the
case where
\begin{IEEEeqnarray}{c}
  \alpha<\alpha_{\textnormal{th}}. 
\end{IEEEeqnarray}

Consider the upper bound in Proposition~\ref{prop:finiteSNR}. We relax
this bound by choosing specific values for $\delta$ and $\mu$
depending on $\vect{p}=(p_1,\ldots, p_{\nt})$ and $\amp$. The relaxed
upper bound will establish the converse to \eqref{eq:asympcap}.

Fix $\amp \ge 1$. For any $\vect{p}=(p_1,\ldots, p_{\nt})$ satisfying
\eqref{eq:m22}, define
\begin{IEEEeqnarray}{c}
  \lambda = \lambda(\vect{p}) \eqdef  \alpha -
  \sum_{k=1}^{\nt} p_k(k-1),
  \label{eq:11}
\end{IEEEeqnarray}
and fix some $0 < \zeta < 1$. We choose
\begin{IEEEeqnarray}{rCl}
  \delta & = &  \log(1+\amp)
  \label{eq:39}
\end{IEEEeqnarray}
and 
\begin{IEEEeqnarray}{rCl}
   \mu =
  \begin{cases}
    \mu^*(\vect{p}) & \textnormal{if } \frac{1}{\amp^{1-\zeta}} <
    \lambda(\vect{p}) < \frac{1}{2}, 
    \\
    \amp^{1-\zeta} & \textnormal{if } \lambda(\vect{p})  \le 
    \frac{1}{\amp^{1-\zeta}},
    \\
    \frac{1}{\amp} & \textnormal{if } \lambda(\vect{p})  \ge
    \frac{1}{2}, 
  \end{cases}
  \label{eq:10}
\end{IEEEeqnarray}
where $\mu^*(\vect{p})$ is the unique positive solution to 
\begin{IEEEeqnarray}{c}
  \label{eq:m60}
  \frac{1}{\mu^*} - \frac{\ope^{-\mu^*}}{1-\ope^{-\mu^*}}
  = \lambda(\vect{p}).
\end{IEEEeqnarray}
Note that in the first case of \eqref{eq:10},
\begin{IEEEeqnarray}{c}
  \frac{1}{\amp^{1-\zeta}} \le \lambda(\vect{p})
  = \frac{1}{\mu^*(\vect{p})} -
  \frac{\ope^{-\mu^*(\vect{p})}}{1-\ope^{-\mu^*(\vect{p})}} 
  \le \frac{1}{\mu^*(\vect{p})} 
  \label{eq:14}
\end{IEEEeqnarray}
and thus
\begin{IEEEeqnarray}{c}
  \label{eq:zeta}
  \mu^*(\vect{p}) \leq \amp^{1-\zeta}.
\end{IEEEeqnarray}
Our choice \eqref{eq:10} thus guarantees in all three cases that
\begin{IEEEeqnarray}{c}
  \mu \le \amp^{1-\zeta},
  \qquad \textnormal{for } \amp \geq 1,
  \label{eq:21}
\end{IEEEeqnarray}
and we can bound
\begin{IEEEeqnarray}{rCl}
  \frac{\mu\sigma}{\amp\sqrt{2\pi}} \sum_{k=1}^{\nt} \frac{p_k}{h_k}
  \left(\ope^{-\frac{\delta^2}{2\sigma^2}}  
    - \ope^{-\frac{(\amp h_k+\delta)^2}{2\sigma^2}}\right)
  & \le & \frac{\sigma}{\amp^{\zeta}\sqrt{2\pi}}
  \sum_{k=1}^{\nt} \frac{1}{h_k}
  \left(\ope^{-\frac{\delta^2}{2\sigma^2}}  
    - \ope^{-\frac{(\amp h_k+\delta)^2}{2\sigma^2}}\right)
  \IEEEeqnarraynumspace
\end{IEEEeqnarray}
and
\begin{IEEEeqnarray}{rCl}
  \sum_{k=1}^{\nt} p_k \log \left( 
    \ope^{\frac{\mu\delta}{\amp h_k}} -
    \ope^{-\mu-\frac{\mu\delta}{\amp h_k}}\right)
  & \le & \sum_{k=1}^{\nt} p_k \log \left( 
    \ope^{\frac{\delta \amp^{-\zeta}}{h_{\nt}}} -
    \ope^{-\mu
      - \frac{\delta\amp^{-\zeta}}{h_{\nt}}}\right)    
  \\
  & = & \log \left( 
    \ope^{\frac{\delta\amp^{-\zeta}}{h_{\nt}}} -
    \ope^{-\mu
      -\frac{\delta\amp^{-\zeta}}{h_{\nt}}}\right),   
\end{IEEEeqnarray}
where we have also used $p_k \le 1$ and $h_k \ge h_{\nt}$.

With the described choices of $\mu, \delta$ and the proposed
relaxations, upper bound \eqref{eq:ub2} becomes
\begin{IEEEeqnarray}{rCl}
  \C_{\trans{\vect{h}},\sigma^2}(\amp,\alpha\amp)
  & \le &\frac{1}{2}\log\frac{\amp^2s_{\nt}^2}{2\pi e\sigma^2}
  + f(\amp) + \sup_{\vect{p}} g(\amp,\vect{p},\mu) 
  \label{eq:m8}
\end{IEEEeqnarray}
with
\begin{IEEEeqnarray}{rCl}
  f(\amp) 
  & \eqdef & 
  \Qf{\frac{\log(1+\amp)}{\sigma}} 
  - \log\left( 1 - 2\Qf{\frac{\log(1+\amp)}{\sigma}}\right) 
  + \frac{\log(1+\amp)}{\sqrt{2\pi}\sigma}
  \ope^{-\frac{\log^2(1+\amp)}{2\sigma^2}}  
  \nonumber\\
  && +\> \frac{\sigma}{\amp^{\zeta}\sqrt{2\pi}} \sum_{k=1}^{\nt}
  \frac{1}{h_k} \left(\ope^{-\frac{\log^2(1+\amp)}{2\sigma^2}}  
    - \ope^{-\frac{(\amp h_k+\log(1+\amp))^2}{2\sigma^2}}\right)
\end{IEEEeqnarray}
and
\begin{IEEEeqnarray}{rCl}
  g(\amp,\vect{p},\mu)
  & \eqdef & - \relD\left(\vect{p}\middle\| \frac{\vect{h}}{s_{\nt}}
  \right)  
  + \log \left(\ope^{\frac{\amp^{-\zeta} \log(1+\amp)}{h_{\nt}}}
    - \ope^{-\mu
      -\frac{\amp^{-\zeta}\log(1+\amp)}{h_{\nt}}}\right) 
  \nonumber\\
  && - \log\mu + \mu \left(\alpha -
    \sum_{k=1}^{\nt} 
    p_k(k-1) \right) .
  \IEEEeqnarraynumspace
\end{IEEEeqnarray}
We note that 
\begin{IEEEeqnarray}{c}
  \label{eq:m9}
  \lim_{\amp\to\infty} f(\amp) = 0.
\end{IEEEeqnarray}

Next, we bound $g\bigl(\amp,\vect{p},\mu\bigr)$ by bounding the
function individually for the three different cases defined
in~\eqref{eq:10}, and then taking the maximum over the three obtained
bounds.  Notice first that when
$\lambda(\vect{p}) = \alpha-\sum_{k=1}^{\nt} p_k(k-1)$ lies in the
open interval $(\amp^{1-\zeta}, 1/2)$,
\begin{IEEEeqnarray}{rCl}
  \IEEEeqnarraymulticol{3}{l}{%
    g(\amp, \vect{p}, \mu) 
  }\nonumber\\*\quad%
  & = & - \relD\left(\vect{p}\middle\| \frac{\vect{h}}{s_{\nt}}
  \right) + \log 
  \left(\ope^{\frac{\amp^{-\zeta} \log(1+\amp)}{h_{\nt}}} - 
    \ope^{-\mu^*(\vect{p})-\frac{\amp^{-\zeta}\log(1+\amp)}{h_{\nt}}}\right) 
  - \log \mu^*(\vect{p})  \nonumber\\
  && +\> \mu^*(\vect{p}) \lambda(\vect{p})
  \\
  & = & -\relD\left(\vect{p}\middle\| \frac{\vect{h}}{s_{\nt}}
  \right) + \log 
  \left(\ope^{\frac{\amp^{-\zeta}\log(1+\amp)}{h_{\nt}}} - 
    \ope^{-\mu^*(\vect{p})-\frac{\amp^{-\zeta}\log(1+\amp)}{h_{\nt}}}\right)  -
  \log \mu^*(\vect{p}) \nonumber\\
  && +\> \mu^*(\vect{p})\left(\frac{1}{\mu^*(\vect{p})} - 
    \frac{\ope^{-\mu^*(\vect{p})}}{1-\ope^{-\mu^*(\vect{p})}} \right)
  \label{eq:m4}
  \\
  & \le & \sup_{\vect{p}\colon \lambda(\vect{p}) \in
    \left(\frac{1}{\amp^{1-\zeta}},\frac{1}{2}\right)} 
  \Biggl\{  
  - \relD\left(\vect{p}\middle\| \frac{\vect{h}}{s_{\nt}} \right)
  - \log \mu^*(\vect{p}) + \mu^*(\vect{p})\left(\frac{1}{\mu^*(\vect{p})}
    - \frac{\ope^{-\mu^*(\vect{p})}}{1-\ope^{-\mu^*(\vect{p})}} \right) 
  \nonumber\\
  && \qquad\qquad\qquad\qquad
  + \log
  \left(\ope^{\amp^{-\zeta}\frac{\log(1+\amp)}{h_{\nt}}} - 
    \ope^{-\mu^*(\vect{p})-\frac{\amp^{-\zeta}\log(1+\amp)}{h_{\nt}}}\right)
  \Biggr\}
  \label{eq:m2}
  \\
  & = & \sup_{\vect{p}\colon \lambda(\vect{p}) \in
    \left(\frac{1}{\amp^{1-\zeta}},\frac{1}{2}\right)} \left\{ 
    - \relD\left(\vect{p}\middle\| \frac{\vect{h}}{s_{\nt}} \right) 
    - \log \frac{\mu^*(\vect{p})}{1-\ope^{-\mu^*(\vect{p})}}
    - \frac{\mu^*(\vect{p})\ope^{-\mu^*(\vect{p})}}{1-\ope^{-\mu^*(\vect{p})}} + 1
    \sizecorr{\frac{\ope^{\frac{2\log(1+\amp)}{\amp^{\zeta} h_{\nt}}} - 
        \ope^{-\mu^*(\vect{p})}}{1-\ope^{-\mu^*(\vect{p})}}} \right.
  \nonumber\\
  && \qquad\qquad\qquad\qquad \left.
        + \log
        \left(\frac{\ope^{\frac{2\amp^{-\zeta}\log(1+\amp)}{h_{\nt}}} 
        - \ope^{-\mu^*(\vect{p})}}{1-\ope^{-\mu^*(\vect{p})}}\right)
  \right\} - \frac{\log(1+\amp)}{\amp^{\zeta} h_{\nt}}
  \label{eq:m3}
  \\
  & \eqdef & g_1(\amp).
\end{IEEEeqnarray}
Here, \eqref{eq:m4} follows from \eqref{eq:m60}; the inequality in
\eqref{eq:m2} holds because
$\lambda(\vect{p})\in \bigl(\amp^{\zeta-1}, 1/2\bigr)$; and
\eqref{eq:m3} follows by rearranging terms.
  
When $\lambda(\vect{p}) \le \amp^{\zeta-1}$, 
\begin{IEEEeqnarray}{rCl}
  \IEEEeqnarraymulticol{3}{l}{%
    g\bigl(\amp, \vect{p}, \mu\bigr)
  }\nonumber\\*\quad%
  & = & - \relD\left(\vect{p}\middle\| \frac{\vect{h}}{s_{\nt}}
  \right) - \log 
  \frac{\amp^{1-\zeta}}{\ope^{\frac{\amp^{-\zeta}\log(1+\amp)}{h_{\nt}}}
    - \ope^{-\amp^{1-\zeta} -
      \frac{\amp^{-\zeta}\log(1+\amp)}{h_{\nt}}}}
  + \amp^{1-\zeta} \lambda(\vect{p}) 
  \\
  & \leq  & - \relD\left(\vect{p}\middle\| \frac{\vect{h}}{s_{\nt}}
  \right)  
  - \log
  \frac{\amp^{1-\zeta}}{\ope^{\frac{\amp^{-\zeta}\log(1+\amp)}{h_{\nt}}}  
    - \ope^{-\amp^{1-\zeta} -
      \frac{\amp^{-\zeta}\log(1+\amp)}{h_{\nt}}}} +1
  \label{eq:m5}
  \\
  & = & \underbrace{\sum_{k=1}^{\nt} p_k\log h_k}_{\le\, \log h_{1}}{}
  + \underbrace{\sum_{k=1}^{\nt} p_k\log \frac{1}{p_k}}_{\le\,\log\nt}{}
  + \log s_{\nt} 
  - \log
  \frac{\amp^{1-\zeta}}{\ope^{\frac{\amp^{-\zeta}\log(1+\amp)}{h_{\nt}}} 
    -  \underbrace{\ope^{-\amp^{1-\zeta} -
        \frac{\amp^{-\zeta}\log(1+\amp)}{h_{\nt}}}}_{\ge\,0}{}}
  \nonumber\\
  && +\> 1
  \label{eq:m6}
  \\
  & \le & - (1-\zeta)\log\amp + \log h_{1} + \log\nt + \log s_{\nt}
  + \frac{\log(1+\amp)}{\amp^{\zeta}h_{\nt}}
  + 1
  \\
  & \eqdef & g_2(\amp),
\end{IEEEeqnarray}
where the inequality in \eqref{eq:m5} follows because
$\lambda(\vect{p}) \le \amp^{\zeta-1}$.

Finally, when $\lambda(\vect{p}) \geq 1/2$,
\begin{IEEEeqnarray}{rCl}
  \IEEEeqnarraymulticol{3}{l}{%
    g(\amp, \vect{p}, \mu)
  }\nonumber\\*\quad%
  & = & - \relD\left(\vect{p}\middle\| \frac{\vect{h}}{s_{\nt}}
  \right) 
  - \log
  \frac{\amp^{-1}}{\ope^{\frac{\amp^{-\zeta}\log(1+\amp)}{h_{\nt}}}
    - \ope^{-\amp^{-1} -\frac{\amp^{-\zeta}\log(1+\amp)}{h_{\nt}}}} 
  + \frac{1}{\amp}\underbrace{\left(
      \alpha - \sum_{k=1}^{\nt} p_k(k-1) \right)}_{\leq \alpha}
  \nonumber\\*
  \\
  & \leq & 
  - \relD\left(\vect{p}\middle\| \frac{\vect{h}}{s_{\nt}} \right) 
  - \log
  \frac{\amp^{-1}}{\ope^{\frac{\amp^{-\zeta}\log(1+\amp)}{h_{\nt}}}  
    - \ope^{-\amp^{-1} -\frac{\amp^{-\zeta}\log(1+\amp)}{h_{\nt}}}}
  + \frac{\alpha}{\amp}
  \\
  & \le & - \inf_{\vect{p}\colon \lambda(\vect{p}) \ge \frac{1}{2}}
  \relD\left(\vect{p}\middle\| \frac{\vect{h}}{s_{\nt}} \right)  
  - \log
  \frac{\amp^{-1}}{\ope^{\frac{\amp^{-\zeta}\log(1+\amp)}{h_{\nt}}}  
    - \ope^{-\amp^{-1} -\frac{\amp^{-\zeta}\log(1+\amp)}{h_{\nt}}}}
  + \frac{\alpha}{\amp} 
  \label{eq:m7}
  \\
  & \eqdef & g_3(\amp).
\end{IEEEeqnarray}
  
We note that depending on the value of $\lambda(\vect{p})$, the
function $g\bigl(\amp, \vect{p}, \mu\bigr)$ is upper-bounded by one of
the three functions $g_1(\amp)$, $g_2(\amp)$, or $g_3(\amp)$. Thus, it
is also upper-bounded by their maximum:
\begin{IEEEeqnarray}{c}
  g(\amp,\vect{p},\mu)
  \le  \max \bigl\{ g_1(\amp), g_2(\amp), g_3(\amp) \bigr\}.
\end{IEEEeqnarray}
We now analyze this maximum when $\amp \to \infty$. Since $g_2(\amp)$
tends to $-\infty$ as $\amp\to \infty$ and since $g_{1}(\amp)$ and
$g_3(\amp)$ are both bounded for $\amp>1$, $g_2(\amp)$ is strictly
smaller than $\max\{g_1(\amp), g_3(\amp)\}$ for $\amp$ large
enough. Moreover,
\begin{IEEEeqnarray}{rCl}
  \lim_{\amp\to\infty} g_3(\amp) 
  & = & - \inf_{\vect{p}\colon \alpha - \sum_{k=1}^{\nt} p_k(k-1) \ge
    \frac{1}{2}} 
  \relD\left(\vect{p} \middle\| \frac{\vect{h}}{s_{\nt}} \right)
  \\
  & = & - \inf_{\vect{p}\colon \alpha - \sum_{k=1}^{\nt} p_k(k-1) =
    \frac{1}{2}} 
  \relD\left(\vect{p} \middle\| \frac{\vect{h}}{s_{\nt}} \right),
  \label{eq:max_3}
\end{IEEEeqnarray}
where the second equality follows because, given
$\alpha<\alpha_{\textnormal{th}}$, an optimal choice of $\vect{p}$
will make full use of the available constraint
\begin{IEEEeqnarray}{c}
  \sum_{k=1}^{\nt} p_k(k-1) \le \alpha - \frac{1}{2}.
\end{IEEEeqnarray}
It remains to investigate the behavior of $g_1(\amp)$ for
$\amp\to\infty$. To that goal define
\begin{IEEEeqnarray}{rCl}
  \tilde{g}_1(\amp,\vect{p})
  & \eqdef & - \relD\left(\vect{p} \middle\| \frac{\vect{h}}{s_{\nt}}
  \right)  
  - \log \frac{\mu^*(\vect{p})}{1-\ope^{-\mu^*(\vect{p})}}
  - \frac{\mu^*(\vect{p})\ope^{-\mu^*(\vect{p})}}{1-\ope^{-\mu^*(\vect{p})}} 
  + 1 \nonumber\\
  && + \log \left(\frac{\ope^{\frac{2\amp^{-\zeta}
          \log(1+\amp)}{h_{\nt}}} 
      - \ope^{-\mu^*(\vect{p})}}{1-\ope^{-\mu^*(\vect{p})}}\right),
\end{IEEEeqnarray}
where $\mu^*(\vect{p})$ is the unique positive solution to
\eqref{eq:m60} (with $\lambda(\vect{p})$ defined in \eqref{eq:11}).

Notice that 
\begin{IEEEeqnarray}{c}
  g_1(\amp) = \sup_{\vect{p}\colon \alpha - \sum_{k=1}^{\nt} p_k(k-1)
    \in \left(\frac{1}{\amp^{1-\zeta}},\frac{1}{2}\right)}
  \tilde{g}_1(\amp,\vect{p})- \frac{\log(1+\amp)}{\amp^{\zeta}
    h_{\nt}}. 
\end{IEEEeqnarray}
Note further that, for a fixed $\vect{p}$,
\begin{IEEEeqnarray}{rCl}
  \Delta(\amp,\vect{p})
  & \eqdef & \tilde{g}_1(\amp,\vect{p}) - \lim_{\amp\to\infty}
  \tilde{g}_1(\amp,\vect{p})
  \\
  & = & \log
  \left(\frac{\ope^{\frac{2\amp^{-\zeta}\log(1+\amp)}{h_{\nt}}} - 
      \ope^{-\mu^*(\vect{p})}}{1-\ope^{-\mu^*(\vect{p})}}\right). 
\end{IEEEeqnarray}
Since $\frac{b-\xi}{1-\xi}$ is increasing in $\xi$ for $b>1$ and
because of \eqref{eq:zeta},
\begin{IEEEeqnarray}{rCl}
  0 \ge \Delta(\amp,\vect{p})
  \ge \log
  \left(\frac{\ope^{\frac{2\amp^{-\zeta}\log(1+\amp)}{h_{\nt}}}
      - \ope^{-\amp^{1-\zeta}}}{1-\ope^{-\amp^{1-\zeta}}}\right)
\end{IEEEeqnarray}
and therefore for any $\vect{p}$:
\begin{IEEEeqnarray}{rCl}
  \bigabs{\Delta(\amp,\vect{p})}
  & \le &  \abs{\log
    \left(\frac{\ope^{\frac{2\amp^{-\zeta}\log(1+\amp)}{h_{\nt}}} 
        - \ope^{-\amp^{1-\zeta}}}{1-\ope^{-\amp^{1-\zeta}}}\right)}
  \stackrel{\amp\to\infty}{\to} \log(1) = 0.
\end{IEEEeqnarray}
This proves that $\tilde{g}_1(\amp,\vect{p})$ converges uniformly as
$\amp \to \infty$ and we are allowed to swap limit and supremum to
obtain
\begin{IEEEeqnarray}{rCl}
  \IEEEeqnarraymulticol{3}{l}{%
    \lim_{\amp\to\infty} g_1(\amp)
  }\nonumber\\*\;%
  & = & \lim_{\amp\to\infty} \left\{
    \sup_{\vect{p}\colon \alpha -
      \sum_{k=1}^{\nt} p_k(k-1) \in
      \left(\frac{1}{\amp^{1-\zeta}},\frac{1}{2}\right)}
    \tilde{g}_1(\amp,\vect{p})  - \frac{\log(1+\amp)}{\amp^{\zeta}
      h_{\nt}} \right\}
  \\
  & = & 
  \sup_{\vect{p}\colon \alpha -
    \sum_{k=1}^{\nt} p_k(k-1) \in (0,\frac{1}{2})}
  \lim_{\amp\to\infty}  \tilde{g}_1(\amp,\vect{p}) 
  \\
  & = & \sup_{\vect{p}\colon \alpha -
    \sum_{k=1}^{\nt} p_k(k-1) \in (0,\frac{1}{2})} \biggl\{
  1 - \log \frac{\mu^*(\vect{p})}{1-\ope^{-\mu^*(\vect{p})}}
  - \frac{\mu^*(\vect{p})\ope^{-\mu^*(\vect{p})}}{1-\ope^{-\mu^*(\vect{p})}}
  - \relD\left(\vect{p}
    \middle\| \frac{\vect{h}}{s_{\nt}}\right) \biggr\}.
  \label{eq:1}
  \IEEEeqnarraynumspace
\end{IEEEeqnarray}
Since for any $\lambda \in \left(0,\frac{1}{2}\right)$,
\begin{IEEEeqnarray}{c}
  \label{eq:m33}
  1- \log \frac{\mu^*}{1-\ope^{-\mu^*}} -
  \frac{\mu^*\ope^{-\mu^*}}{1-\ope^{-\mu^*}} 
  \ge 0 
\end{IEEEeqnarray}
and since
\begin{IEEEeqnarray}{c}
  \lim_{\lambda\to 0} \left\{
    1- \log \frac{\mu^*}{1-\ope^{-\mu^*}} -
    \frac{\mu^*\ope^{-\mu^*}}{1-\ope^{-\mu^*}} 
  \right\} = 0,
\end{IEEEeqnarray}
we conclude by comparing \eqref{eq:max_3} and \eqref{eq:1} that for
sufficiently large values of $\amp$, $g_1(\amp) \geq g_3(\amp)$, and
thus
\begin{IEEEeqnarray}{rCl}
  \IEEEeqnarraymulticol{3}{l}{%
    \lim_{\amp \to \infty} \sup_{\vect{p}} g(\amp, \vect{p}, \mu) 
  }\nonumber\\*\;\;%
  & \leq & \lim_{\amp \to \infty} g_1(\amp)
  \\
  & = & \sup_{\vect{p}\colon \alpha -
    \sum_{k=1}^{\nt} p_k(k-1) \in (0,\frac{1}{2})} \biggl\{
  1 - \log \frac{\mu^*(\vect{p})}{1-\ope^{-\mu^*(\vect{p})}}
  - \frac{\mu^*(\vect{p})\ope^{-\mu^*(\vect{p})}}{1-\ope^{-\mu^*(\vect{p})}}
  - \relD\left(\vect{p}
    \middle\| \frac{\vect{h}}{s_{\nt}}\right) \biggr\}
  \IEEEeqnarraynumspace
  \\
  & = & \sup_{\lambda\in\left(
      \max\{0,\frac{1}{2}+\alpha-\alpha_{\textnormal{th}}\}, 
      \min\{\frac{1}{2},\alpha\}\right)}
  \biggl\{
  1 - \log \frac{\mu(\lambda)}{1-\ope^{-\mu(\lambda)}}
  - \frac{\mu(\lambda)\ope^{-\mu(\lambda)}}{1-\ope^{-\mu(\lambda)}} 
  \nonumber\\
  && \qquad\qquad\qquad\qquad\qquad\qquad\quad
  - \inf_{\vect{p}\colon \alpha -
    \sum_{k=1}^{\nt}  p_k(k-1) = \lambda}     \relD\left(\vect{p}
    \middle\| \frac{\vect{h}}{s_{\nt}} \right)\biggr\},
  \label{eq:4}
\end{IEEEeqnarray}
where in \eqref{eq:4} $\mu(\lambda)$ is the unique positive solution
to \eqref{eq:8}. Note that for the re-parametrizing in \eqref{eq:4} we
have defined
\begin{IEEEeqnarray}{c}
  \lambda = \alpha -  \sum_{k=1}^{\nt}  p_k(k-1) 
\end{IEEEeqnarray}
and then used the same argumentation as given at the end of
Section~\ref{sec:finite-snr-lower-bound} to restrict the required
range of $\lambda$. By \cite[Probl.~12.2]{coverthomas06_1} it then
follows that the infimum is achieved for the $\vect{p}$ given in
\eqref{eq:pk}. 

Combining \eqref{eq:4} with \eqref{eq:m8} and \eqref{eq:m9} then
proves the proposition.

\section{Concluding Remarks}
\label{sec:concluding-remarks}

In this paper we present upper and lower bounds on the capacity of a
multiple-input and single-output (MISO) free-space optical intensity
channel with signal-independent additive Gaussian noise and with both
a peak- and an average-power constraint on the input. Asymptotically,
when both peak and average power tend either to infinity or to zero
(with their ratio held fixed), we succeed in specifying the capacity
exactly.

At low SNR, a good input vector $\vect{X}$ maximizes the variance of
$\trans{\vect{h}}\vect{X}$ under the given power constraints. This is
achieved by $\vect{X}$ having only entries of $\amp$ and $0$, i.e., by
each LED sending either full or no power.

At high SNR, a good input vector $\vect{X}$ maximizes the differential
entropy $\hh(\trans{\vect{h}}\vect{X})$.  For the case of only an
average-power constraint or only a peak-power constraint (or both a
peak- and an average-power constraint but with the latter being
sufficiently loose), this is relatively straightforward. For the
general situation of both a peak- and an average-power constraint,
maximizing $\hh(\trans{\vect{h}}\vect{X})$ is more involved. The
optimal input can be found based on two insights. First, in order to
reach a certain range of amplitude levels
$\trans{\vect{h}}\vect{X} \in (s_{k-1}\amp,s_k\amp]$, it is most
energy-efficient to set all LEDs with strong channel gains to the
maximum level, $X_j= \amp$, $j=1, \ldots, k-1$; to switch the weaker
LEDs off, $X_j=0$, $j=k+1, \ldots, \nt$; and to exclusively use $X_k$
to signal.  Second, conditional on a given range
$(s_{k-1}\amp,s_k\amp]$, $X_k$ should have a truncated exponential
distribution in order to maximize the conditional differential entropy
under the given power constraints. It then only remains to optimize
over the probability masses assigned to each of the different
amplitude ranges and the parameters of the truncated
exponentials. Note that this optimization characterizes an implicit
trade-off: higher probabilities on the higher amplitude ranges will
increase the effectively used total range of
$\trans{\vect{h}}\vect{X}$, but at the cost of using more power for
the LEDs that are set deterministically to $\amp$.

Our lower bound in Proposition~\ref{prop:lower} is based on such a
choice of input distribution. Our upper bound in
Proposition~\ref{prop:finiteSNR} and its asymptotic analysis in
Section~\ref{sec:asymptotic} are based on the same intuition, but
borrow from known upper bounds on the SISO capacity. Alternatively,
one can also derive a new upper bound using the duality-based bounding
technique \cite{lapidothmoser03_3}, as we outlined in
\cite{moserwangwigger17_1app}. Such a bound, however, is much harder
to prove.

A close look at the results in \cite{mosermylonakiswangwigger17_1}
confirms that also for the MIMO optical intensity channel when the
channel matrix $\mat{H}$ has full column rank, the high-SNR asymptotic
capacity is given by the maximum differential entropy of
$\mat{H}\vect{X}$ minus that of the noise vector.  With the current
work and \cite{mosermylonakiswangwigger17_1}, the only MIMO optical
intensity channels whose high-SNR asymptotic capacities are not yet
known are those with more than one receive antennas (photodetectors),
and with channel matrices that do not have full column rank. It is
natural to conjecture that, for those channels, the high-SNR
asymptotic capacity is again given by the maximum of
$\hh(\mat{H}\vect{X})$ minus the differential entropy of the noise.

\section*{Acknowledgments}

We gratefully acknowledge helpful discussions with Tobias Koch and
Christoph Pfister. In particular, Tobi's suggestion helped us simplify
Proposition~\ref{prop:finiteSNR} and its proof.

\appendix

\section{Proof of Lemma~\ref{lem:maximum_variance}}
\label{app:variance}

The variance of $\Xs$ can be decomposed as
\begin{IEEEeqnarray}{rCl}
  \E{\bigl(\Xs-\eE{\Xs}\bigr)^2}
  & = & \sum_{k=1}^{\nt} h_k^2 \E{(X_k-\E{X_k})^2} 
  + \sum_{\substack{i,j=1\\i\neq j}}^{\nt} h_i h_j \bigl( \E{X_i X_j} -
    \E{X_i}\E{X_j}\bigr). 
  \label{eq:ligong139}
  \IEEEeqnarraynumspace
\end{IEEEeqnarray}
Let us fix the joint distribution on $(X_1,\ldots,X_{\nt-1})$, and fix
with probability one the conditional mean
$\Econd{X_{\nt}}{X_1,\ldots,X_{\nt-1}}$. These determine the consumed
average input power, as well as every summand on the RHS of
\eqref{eq:ligong139} except $\E{(X_{\nt} - \E{X_{\nt}})^2}$. For any
choice above, the value of $\E{(X_{\nt} - \E{X_{\nt}})^2}$ is
maximized by $X_{\nt}$ taking value only in the set $\{0,\amp\}$. We
hence conclude that, to maximize the variance \eqref{eq:ligong139}
subject to a constraint on average input power, it is optimal to
restrict $X_{\nt}$ to taking value only in $\{0,\amp\}$. Repeating
this argument, we conclude that every $X_k$, $k=1,\ldots,\nt$, should
take value only in $\{0,\amp\}$.

Next, using the same argument as in Lemma~\ref{lem:energy_efficient},
we know that it is optimal to consider joint distributions as follows:
for each $k\in\{0,\ldots, \nt\}$, with probability $q_k$
\begin{IEEEeqnarray}{c}
  \label{eq:choice}
  X_1 = \cdots = X_k = \amp \qquad \textnormal{ and } \qquad
  X_{k+1} = \cdots = X_{\nt} = 0. 
\end{IEEEeqnarray}
Such a choice produces an $\Xs$ that takes value only in
\eqref{eq:nt+1}. This proves Part~1 of the lemma. Further, this choice
of inputs consumes an average power of $\sum_{k=0}^{\nt} q_k k$.  The
condition for a probability vector $q_0, \ldots, q_{\nt}$ to be valid
is thus
\begin{IEEEeqnarray}{c}
  \sum_{k=0}^{\nt} q_k  k \leq \alpha.
\end{IEEEeqnarray} 
Part~2 of the lemma is then proven by noting that with the choice in
\eqref{eq:choice}, the variance of $\Xs$ is
\begin{IEEEeqnarray}{rCl}
  \E{\bigl(\Xs-\eE{\Xs}\bigr)^2}
  & = &	\E{\Xs^2} - \bigl(\eE{\Xs}\bigr)^2
  \\
  & = & \sum_{k=1}^{\nt} q_k \amp^2 s_k^2 - \left(
    \sum_{k=1}^{\nt} q_k \amp s_k  \right)^2. 
\end{IEEEeqnarray}



\end{document}